\documentclass[prb,superscriptaddress,twocolumn,amsmath,amssymb,floatfix]{revtex4-1}
\usepackage{latexsym}
\usepackage{dcolumn}
\usepackage[dvips]{graphicx}
\usepackage{amssymb}
\usepackage{graphics}
\usepackage{amsmath}
\usepackage{epsf}
\usepackage{bigstrut}
\usepackage{multirow}
\usepackage[english]{babel}

\begin{document}
\title{Universal density scaling of disorder-limited low-temperature
  conductivity in high-mobility two-dimensional systems}
\author{S. Das Sarma}
\affiliation{Condensed Matter Theory Center, Department of Physics, 
	 University of Maryland,
	 College Park, Maryland  20742-4111}
\author{E. H. Hwang}
\affiliation{Condensed Matter Theory Center, Department of Physics, 
	 University of Maryland,
	 College Park, Maryland  20742-4111}
\affiliation{SKKU Advanced Institute of Nanotechnology and Department
  of Physics, Sungkyunkwan
  University, Suwon 440-746, Korea }

\date{\today}

\begin{abstract}
We theoretically consider the carrier density dependence of
low-temperature electrical conductivity in high-quality and
low-disorder two-dimensional (2D) ``metallic'' electronic systems such
as 2D GaAs electron or hole quantum wells or doped/gated
graphene. Taking into account resistive scattering by Coulomb disorder
arising from quenched random charged impurities in the environment, we
show that the 2D conductivity $\sigma(n)$ varies as $\sigma \sim
n^{\beta(n)}$ as a function of the 2D carrier density $n$ where the
exponent $\beta(n)$ is a smooth, but non-monotonic, function of density
$n$ with possible nontrivial extrema. In particular, the density
scaling exponent $\beta(n)$ depends qualitatively on whether the
Coulomb disorder arises primarily from remote or background charged
impurities or short-range disorder, and can, in principle, be used to
characterize the nature of the dominant background disorder. A
specific important prediction of the theory is that for resistive
scattering by remote charged impurities, the exponent $\beta$ can
reach a value as large as 2.7 for $k_F d \sim 1$, where $k_F \sim
\sqrt{n}$ is the 2D Fermi wave vector and $d$ is the separation of the
remote impurities from the 2D layer. Such an exponent $\beta$ ($>5/2$)
is surprising because unscreened Coulomb scattering by remote
impurities gives a limiting theoretical scaling exponent of $\beta =
5/2$, and n\"{a}ively one would expect $\beta(n) \le 5/2$ for all
densities since unscreened Coulomb scattering should nominally be the
situation bounding the resistive scattering from above. We find
numerically and show theoretically that the maximum value of $\alpha$
($\beta$), the mobility (conductivity) exponent, for 2D semiconductor
quantum wells is around 1.7 (2.7) for all values of $d$ (and for both
electrons and holes) with the maximum $\alpha$ occurring around $k_F d
\sim 1$. We discuss experimental scenarios for the verification of our
theory. 
\end{abstract}

\maketitle

\section{introduction}

It is well-known that carrier scattering by background quenched
disorder arising from random impurities and defects limits the $T=0$
(i.e., low-temperature) residual conductivity of a metal or doped
semiconductor. This ``residual resistivity'', i.e., the extrapolated
$T=0$ value of the electrical resistivity obtained from the measured
low-temperature transport data, provides information about the
extrinsic disorder in the material limiting transport properties. In
3D metals, this residual resistivity is not of much intrinsic
fundamental interest except for the fact that very pure defect-free
single crystals, where the background disorder is presumably very low,
could have extremely low residual resistivity leading to very long
electron mean free paths. In particular, the residual resistivity of
different samples (with different levels of sample purity, i.e.,
different impurity or defect content) of the same metal could differ
by orders of magnitude, and a metallic resistivity uniquely
defining a particular metal (e.g. Cu, Al, or Ag) is only meaningful at
higher temperatures ($\agt 100$ K) where phonon scattering dominates
the electrical resistivity over impurity/defect/disorder
scattering. At low temperatures, each metallic sample would have a
unique resistivity reflecting its specific impurity content signature,
and as such characterizing a metal by a unique resistivity is useless
at low temperatures (i.e., each sample of the same metal would have
different resistivity at $T=0$). In fact, the low-temperature
resistivity of a particular sample typically depends on the
preparation history of the sample, and annealing to room temperatures
(where all samples of a particular metal do have the same resistivity)
and then cooling down to low temperatures could substantially modify
the sample resistivity as the impurity/disorder configuration could
change due to annealing. Theoretical statements \cite{one} about
residual resistivity (or equivalently, low temperature resistivity) of
3D metals thus focus on first principles
calculations of the quantitative aspects of the local disorder
potential arising from various defects or disorder in the metal (where
the specific types of defect or impurity have to be specifically
assumed), and then estimating the resistivity arising from various
postulated disorder in terms of resistivity due to some specified
impurity or defect content on an atomic percentage basis. Typical
values of calculated residual resistivity for most simple metals fall
in the $\sim 0.1-1$ $\mu \Omega$cm range per atomic percentage of
impurities, and thus a metal with 99.999\% (e.g., Cu) purity could
have an extremely low residual resistivity of $10^{-10}$ $\Omega$cm,
leading to elastic mean free paths $\sim 1$ cm although the typical
phonon-limited room-temperature mean free paths in most metals are
only $1-10$ \AA \; with the room-temperature phonon-limited
resistivity of most metals being around $\sim \mu\Omega$cm.

Thus, theoretical studies \cite{one,two} of disorder-limited residual
resistivity of metals focus entirely on the quantitative modeling of
various defect scattering in the metal using detailed
materials-specific band structure and Boltzmann transport theories. No
systematic dependence of the residual resistivity of metals on
various metallicity parameters (e.g., the Wigner-Seitz radius $r_s$ or
lattice constant or Fermi energy) can in general be discussed
qualitatively, and there is no metallic density scaling of
conductivity that one can speak of since the carrier density cannot be
tuned by an external gate voltage in metals (as it can be in 2D
semiconductor systems or in graphene). In principle, there is only a
modest variation in $r_s$ ($\sim 2-6$) among different 3D metals
(since $r_s \sim n^{-1/3}$ where $n$ is the effective metallic free
electron density), but the band structure variation in going from a
metal to another would typically swamp any systematic $r_s$-dependence
of the residual resistivity. Thus, the only systematic trend of the
residual resistivity in metals which is discussed in the literature
\cite{one, two} is the dependence of the $T=0$ impurity-limited
resistivity of a particular metal on the atomic type (e.g. atomic
number) of the various impurities or on the type of defects causing
the resistive scattering. It is basically a subject of quantitative
details based on serious numerical calculations.

In the current work we are interested in discussing qualitatively the
impurity-limited `metallic' residual (i.e. $T=0$) resistivity as a
function of the electron density of the metal. In particular, we want
to understand the density-dependence of the $T=0$ metallic resistivity
assuming that the metallic density can be tuned continuously keeping
all other parameters (e.g., disorder, band structure) fixed. Of
course, such a situation is practically impossible in 3D metals since
the electron density is fixed in each metal and cannot be tuned at
all. But, as is well-known, such a tunable carrier density situation
is routine in 2D ``metallic'' systems such as Si MOSFETs, gated
semiconductor heterostructures and quantum wells, and gated
graphene. In principle, a variable carrier density can be achieved in
3D doped semiconductor systems by changing the doping level (but
keeping the same dopant element) in different samples, but this is not
ideal for our purpose of a qualitative understanding of the density
scaling of the metallic $T=0$ saturated resistivity since there is
always the possibility of unknown sample to sample variation beyond
just the carrier density variation since doping level cannot be tuned
continuously in a single sample as can be done in 2D systems. We do,
however, briefly consider the density
scaling properties of 3D impurity-limited saturated resistivity for
the sake of completeness although most of the current work focuses on
2D semiconductor structures (and graphene) where the carrier density
can be easily tuned in the same sample by applying an external gate
voltage, and thus the density-dependent conductivity is a meaningful
concept in 2D ``effective metallic'' systems existing in semiconductor
structures and graphene.

Obviously, the $T=0$ conductivity ($\equiv \rho^{-1}$, where $\rho$ is
resistivity) would manifest different density dependence for different
types of background disorder, i.e., different types of
impurity-electron interaction. The main resistive disorder scattering
in 
relatively pure 3D metals is due to defects, vacancies, and impurities
which scatter primarily through the short-range (essentially)
$\delta$-function type scattering potential (although there are
extended defects which could produce anisotropic longer-range disorder
potential). Disorder scattering by $\delta$-function type point
scatterers gives rise to rather uninteresting density scaling of
electrical conductivity, again making 3D metals unsuitable for
studying the density dependence of conductivity. Our main interest in
this work is to obtain the density scaling of 2D $T=0$ ``metallic''
conductivity arising from Coulomb disorder induced by random quenched
charged impurity centers in the environment. Quenched Coulomb disorder
is the dominant extrinsic resistive scattering mechanism in all
semiconductor systems (2D or 3D) at low temperatures since doping by
impurities 
is essential in inducing carriers in a semiconductor (whereas metals,
by definition, have free carriers at the Fermi level even at $T=0$),
leading to the inevitable existence of background Coulomb
disorder. Thus, a key distinction between low-temperature transport in
metals and semiconductors is that short-range disorder (arising from
very strong metallic screening) dominates metallic transport whereas
long-range Coulombic disorder (arising from random localized charged
impurities) dominates transport in (both 2D and 3D) doped
semiconductors. The density dependent conductivity in doped
semiconductor structures (or graphene) arises from the carrier
screening of background Coulombic disorder which depends sensitively
on the dimensionless ratio $q_{TF}/2k_F$ where $q_{TF}$ and $k_F$ are
respectively the (density dependent) screening 
wave vector and Fermi wave vector of the system. The variation in the
carrier screening properties as a function of the carrier density
eventually leads to the density dependence of the resistivity. 

Of course in doped semiconductors, one must worry about the additional
complications of disorder-induced Anderson localization and/or
low-temperature carrier freezeout and/or possible percolation
transition associated with charged impurity-induced charge puddle
formation. In the current work, we ignore all of these complications
uncritically, focusing primarily on extremely high-quality
modulation-doped 2D GaAs quantum well structures (and high-quality
suspended graphene) where these complicating circumstances are absent
down to very low carrier densities and very low temperatures. Our
theoretical results presented in this paper, based on Drude-Boltzmann
semiclassical transport theory, should apply to experimental systems
above carrier densities (and temperatures) where localization (and
related effects) become operational. 
The neglect of carrier localization and freezeout is not a
particularly significant issue for our 2D theory and results because
high-quality 2D semiconductor systems (and graphene) are not
susceptible to these problems except perhaps at extremely low
temperatures and carrier densities of little practical or experimental
interest. On the other hand, our low-temperature and low-density
transport results for 3D doped semiconductors are given here purely for
the purpose of completeness and comparison with 2D results since 3D
doped semiconductors typically become insulating at low carrier
density (as well as low temperature) because of localization and
carrier freezeout. Our focus and most of our presented results, as the
title of our article clearly indicates, are on low-temperature density
dependent metallic transport properties of very high-quality 2D
systems where the 
concept of density scaling of conductivity is both theoretically and
experimentally meaningful.

One may wonder about the fact that the scaling theory of localization
predicts that all 2D systems are strictly speaking Anderson
insulators, and have, in principle, zero conductivity at $T=0$ for
infinite samples at all carrier densities. (This is theoretically true
even for graphene when disorder induced intervalley carrier scattering
is taken into account.) For our purpose in the current work, where
we are specifically interested in very high-mobility 2D structures
with very low background disorder, the scaling localization is a
non-issue because (1) the samples are finite in size and the
temperature, although low, is still finite; and (2) more importantly,
the elastic mean free path is extremely long, making the effective 2D
localization length much larger than the system size (or the phase
breaking length, whichever is shorter). Thus, we are specifically
considering 
the density dependence of the semiclassical part of the 2D resistivity
in the situation where the semiclassical resistivity is much larger
than the quantum (or weak localization) contribution to the
resistivity. This limit is generic in high mobility 2D GaAs
semiconductor systems and in graphene, and therefore our work is of
wide validity. Thus we are strictly speaking considering the density
dependence of the zeroth order semiclassical conductivity in the
situation where the quantum contribution to the resistivity is
negligible.

We emphasize that the density dependence (or scaling) of electrical
transport properties is rarely discussed in the theoretical literature
mainly because (as discussed above) such a density dependence of
electrical conductivity is neither relevant nor interesting in 3D
metals. Indeed, it is universally believed that the only  aspect of
electrical conductivity where studying the carrier density dependence
is a meaningful endeavor is in the study of density-tuned Anderson
localization as a ($T=0$) quantum phase transition, exactly the
physical phenomenon we are excluding in the current work where we
focus entirely on the Drude-Boltzmann part of the conductivity in high
quality samples assuming quantum localization effects to be
negligible. Thus, the density scaling of our theory is not connected
with quantum criticality at all, but with the behavior of the density
dependence of the background disorder arising from nontrivial
screening properties. In contrast to the density dependence which is
rarely discussed in the literature except in the context of
metal-insulator transition, the temperature dependence of transport
properties has been extensively discussed for electronic materials
(including many experimental and theoretical papers in 2D systems
\cite{three,four,five,six,seven}) because the temperature dependent
electrical resistivity generically contains qualitative information
about the underlying resistive scattering mechanism. For example,
a linear-in-T higher temperature metallic resistivity is the hallmark
of acoustic phonon scattering in both metals and semiconductors (and
in both 2D or 3D systems) whereas acoustic phonons typically lead to
strongly suppressed high power laws ($\sim T^{\eta}$ with $\eta \approx
4-7$ depending on the details) in the (2D or 3D) metallic resistivity
below a system-dependent Bloch-Gr\"{u}neisen temperature $T_{BG}$. Our
current work focuses entirely on temperatures well below the
Bloch-Gr\"{u}neisen temperature ($T \ll T_{BG}$) so that phonons are
completely ineffective in limiting the electrical
conductivity. Optical phonons, which are often relevant in
semiconductor transport at higher temperatures and indeed limit the
room-temperature resistivity of 2D GaAs-based semiconductor structures
\cite{eight} and are exponentially suppressed as $e^{-\hbar
  \omega_{LO}/k_B T}$ at low temperatures (where $\hbar \omega_{LO}
\sim 450$ K in GaAs), are completely irrelevant for our
low-temperature transport considerations. 

It is well-known that high-quality low-density 2D semiconductor
systems (as well as graphene) often manifest a strong temperature
dependence in its low-temperature resistivity
\cite{three,four,five,six,seven}, which arises from the strong
temperature dependence of the screened Coulomb disorder \cite{nine} at
low carrier densities. Again, our theoretical work at $T=0$ is
completely free from any temperature-induced complications in the
resistivity since our interest is in understanding the density
dependence of the $T=0$ resistivity/conductivity, which can be
obtained from the low-temperature experimental data by extrapolation
to $T=0$ or by sitting always at a constant very low temperature
(e.g., 50 mK) in obtaining the density dependence of the transport
properties. In any case, the question we are interested in is
perfectly well-defined
as a matter of principle: How does the $T=0$ Drude-Boltzmann
semiclassical conductivity of a 2D (or 3D) ``metallic'' system vary as
a function of its carrier density in the presence of background
screened Coulomb disorder being the main resistive scattering
mechanism?

In the rest of this article, we provide a detailed theoretical answer
to the question posed in the last sentence above.


\section{background}

The $T=0$ conductivity $\sigma(n)$ of a 2D metallic system is
typically written as \cite{nine,ten}
\begin{equation}
\sigma(n) = \frac{e^2 v_F^2}{2} D(E_F)\tau(E_F),
\label{eq:eq1}
\end{equation}
in the Boltzmann theory, where $E_F$, $v_F$ are respectively the Fermi
energy and the Fermi velocity, $D(E_F)$ is the carrier density of states at
the Fermi surface, and $\tau$ is the so-called relaxation time (or the
scattering time) for the relevant resistive scattering mechanism. (The
factor 2 in the denominator of Eq.~(\ref{eq:eq1}) is replaced by 3
for 3D systems.) It is assumed that $v_F$, $E_F$, and $D$ are known as a
function of the carrier density $n$ from the relevant band structure
information (and we will consider only the parabolic and the linear
band approximation with the linear approximation used for obtaining
our graphene
transport results). The whole theory then boils down to a calculation
of the transport relaxation time $\tau$ at the Fermi surface using the
appropriate microscopic scattering mechanism (which we describe in
details in section III of our paper).

For parabolic bands with $E({\bf k}) = \hbar^2 k^2/2 m$ we can write
$E_F=mv_F^2/2$, where $m$ is the carrier effective mass, and
Eq.~(\ref{eq:eq1}) simply reduces to the celebrated Drude-Boltzmann
transport formula:
\begin{equation}
\sigma(n) = \frac{n e^2 \tau}{m},
\label{eq:eq2}
\end{equation}
where $\tau\equiv \tau(E_F)$ and $n$ is the relevant (2D or 3D)
carrier density. For graphene, Eq.~(\ref{eq:eq2}) requires a slight
modification \cite{nine} which we will discuss later when we come to
describing 
our graphene results. From this point on, unless otherwise stated, our
explicit equations and formula are given for parabolic band 2D systems
--- we will consider the very special linear band case of graphene
separately at the appropriate juncture. Most of our results focus on
2D semiconductor systems, specifically 2D GaAs electron or hole
quantum wells where the parabolic approximation applies well. We will
point out the corresponding 3D parabolic band analytic results as
appropriate, concentrating on presenting the equations and formulas
mainly for 2D quantum well systems since most of our presented results
are for 2D semiconductor systems, where the corresponding experiments
are feasible.

We note that one can, instead of discussing the conductivity $\sigma$,
equivalently discuss the relaxation
time $\tau$ or the resistivity $\rho = \sigma^{-1}$ or the mobility
$\mu$, which is defined as 
\begin{equation}
\mu = \frac{e \tau}{m} = \frac{\sigma}{ne}.
\end{equation}
Obtaining the density dependent conductivity $\sigma(n)$ now becomes
equivalent to obtaining the density dependence of $\tau$ or $\mu$
since $\sigma \propto n \mu$ (or $n\tau$). We write:
\begin{equation}
\tau \propto n^{\alpha}, \;\; {\rm i.e.}, \;\; \mu \propto n^{\alpha},
\end{equation}
and therefore 
\begin{equation}
\sigma \sim n^{\beta = \alpha + 1}.
\end{equation}
In general, we can define the conductivity exponent $\beta$, or
equivalently the mobility exponent $\alpha$ through the relation:
\begin{equation}
\alpha(n) = \frac{d \ln \mu}{d \ln n},
\end{equation}
and
\begin{equation}
\beta(n) = \alpha(n) + 1 = \frac{d \ln \sigma}{d \ln n}.
\end{equation}
We note that, in general, the mobility exponent $\alpha \equiv
\alpha_{\mu}$ and the relaxation rate exponent $\alpha_{\tau}$ may be
unequal (e.g., graphene) except that for parabolic bands we have
$\alpha_{\mu} = 
\alpha_{\tau} = \alpha$. But the relation $\beta = \alpha_{\mu} +
1=\alpha + 1$ always applies. The main goal of the current work is to
calculate the exponent $\alpha(n)$ for various specified disorder
mechanism, and discuss/contrast how the density scaling exponent
depends on the type of disorder controlling the resistive scattering
mechanism.

We will obtain $\alpha(n)$ analytically in various limiting situations
in our theoretical study. But the main technique would be to obtain
$\mu(n)$ or $\sigma(n)$ numerically from the Boltzmann theory and then
to calculate $\alpha(n)$ or $\beta(n)$. Experimentally, the way to
estimate the exponent $\alpha(n)$ is to plot the measured
low-temperature $\mu(n)$ against $n$, and then to estimate $\alpha(n)$
by carrying out the logarithmic differentiation. In general,
$\alpha(n)$ will depend strongly on the background disorder, and will
vary smoothly as a function of carrier density as different scattering
mechanisms are operational in different density regimes and as the
background Coulomb disorder is screened differently at different
carrier density through the variation in $q_{TF}/2k_F$. An important
surprising finding of our theory is an interesting nonmonotonic
variation in the scaling exponent $\alpha(n)$ as a function of carrier
density.

It is important to point out at this stage that graphene, because of
its linear dispersion with a constant Fermi velocity ($v_F \equiv
v_0$), does not obey \cite{nine} the exponent scaling relation $\beta =
\alpha_{\tau} + 1$ connecting the density scaling between conductivity
($\beta$) and relaxation rate ($\alpha_{\tau}$). Instead, for graphene
\cite{nine}, we have $\beta = \alpha_{\tau} + 1/2$ if $\alpha_{\tau}$
is defined by $\tau^{-1} \sim n^{\alpha_{\tau}}$ which follows from
the constancy of $v_F$ and the fact that $D(E_F) \propto E_F \propto
\sqrt{n}$ in graphene. Putting $D(E_F) \propto k_F \propto \sqrt{n}$
in Eq.~(\ref{eq:eq17}), we get for graphene $\sigma(n) \sim \sqrt{n}
\tau(n)$, i.e., $\beta = \alpha_{\tau}+1/2$. Thus, whereas in ordinary
parabolic systems the exponent $\alpha$ ($\equiv \alpha_{\tau} \equiv
\alpha_{\mu}$) is the same for both mobility ($\mu$) and the
relaxation rate ($\tau^{-1}$), in graphene, by virtue of its linear
band dispersion, $\alpha=\alpha_{\mu} = \alpha_{\tau}-1/2$, but the
relationship between the conductivity exponent ($\beta$) and the
mobility exponent ($\alpha$) is still given by $\beta = \alpha + 1$
since by definition $\sigma = ne \mu$.

\section{Model and Theory}

The central quantity to obtain theoretically in the semiclassical
Boltzmann transport theory in the relaxation time $\tau$ or
equivalently the relaxation rate $\tau^{-1}$, which is given by the
following expression for 2D systems, within the leading order Born
approximation, for carrier scattering at $T=0$ by disorder
\cite{nine,ten}: 
\begin{eqnarray}
\frac{1}{\tau} = \frac{2\pi}{\hbar} \sum_{\gamma} & &\int
N_i^{(\gamma)}(z) dz \int 
\frac{d^2 k'}{(2\pi)^2} \left |{V^{(\gamma)}_{\bf k-k'}(z)}
\right |^2 \nonumber \\
&\times & (1-\cos \theta_{\bf kk'}) \delta [E({\bf
  k})-E({\bf k'})].
\label{eq:eq7}
\end{eqnarray}
Here $N_i^{(\gamma)}(z)$ is the 3D density of random impurities of the
$\gamma$-th type (in general, there could be several different types of
impurities present in the system: near and far, 2D or 3D, long- or
short-range) with $z$
being the direction perpendicular to the plane of 2D system (which
lies in the $x$-$y$ plane); $V_{\bf q}(z)$ is the electron-impurity
interaction (in the 2D wave vector space defined by {\bf q}).
${\bf k, k'}$ are the incoming and outgoing 2D carrier wave vectors
involved in the scattering process with a scattering angle
$\theta_{\bf kk'}$ between them and ${\bf k-k'}$ being the scattering
wave vector; $E({\bf k}) = \hbar^2 k^2/2m$ is the carrier energy. Note
that our disorder model assumes a random distribution of impurities in
the 2D $x$-$y$ plane although it is easy to include correlations in
the 2D impurity distribution if experiment indicates the importance of
such correlations.

Once the scattering potential $V_{\bf q}(z)$ is defined, the problem
of calculating the 2D conductivity becomes simply a question of
evaluating the 4-dimensional integral given in
Eq.~(\ref{eq:eq7}). Note that we are restricting to the $T=0$ case
(or to low temperatures), otherwise a thermal average would be
required in defining $\tau^{-1}$ necessitating a 5-dimensional
integration. We note that although Eq.~(\ref{eq:eq7}) applies only
to 2D systems, a very slight modification gives us the corresponding
expressions for 3D systems and graphene, which we do not show here. We
note here that for graphene there is a well-known additional form
factor of ($1+\cos\theta_{\bf kk'}$) inside the integral in
Eq.~(\ref{eq:eq7}) arising from chirality \cite{nine}.

It is worthwhile to point out here that the theoretical idea of a
meaningful universal density scaling behavior of conductivity applies
as a matter of principle only when the resistive scattering is
dominated by a particular disorder mechanism. If there are many
different types of disorder (i.e., several $\gamma$-types)
contributing equivalently to the resistivity, then the net resistivity
will be given by the Matthiessen's rule: $\tau^{-1}(n) = \sum_{\gamma}
\tau_{\gamma}^{-1}(n)$, and $\rho(n) = \sum_{\gamma}\rho_{\gamma}(n)$,
i.e., $ \sigma^{-1}(n) = \sum_{\gamma} \sigma^{-1}_{\gamma}(n)$, 
where $\rho_{\gamma}$, $\sigma_{\gamma}$, $\tau_{\gamma}$ are
respectively the resistivity, conductivity, scattering time for the
$\gamma$-th type of disorder arising from $N_i^{(\gamma)}(z)$ in
Eq.~(\ref{eq:eq7}). In such a situation, unless one particular type of
disorder (i.e., one specific $\gamma$) dominates scattering, the
resulting density dependence of the total $\sigma$ (or $\mu$) will
manifest complex crossover behavior arising from the combination of
all different scattering processes contributing with different
strength, and there will not be only universal density dependent
scaling behavior of $\sigma(n)$ or $\mu(n)$. To exemplify this
important point, we consider a strict 2D electron gas being scattered
by three different types of disorder (i.e., $\gamma=1$, 2, 3) given by
remote random charged impurities at a distance $d$ from the 2D system,
background charged impurities at the 2D layer, and zero-range disorder
in the layer, each with their respective conductivity exponent
$\beta_{\gamma}$ with $\gamma = 1$, 2, 3 respectively. We can formally
write:
\begin{eqnarray}
\sigma^{-1}(n) &  = & \sigma^{-1}_1 + \sigma^{-1}_2 + \sigma^{-1}_3
\nonumber \\
              & = & A_1n^{-\beta_1} + A_2 n^{-\beta_2} + A_3 n^{-\beta_3},
\end{eqnarray}
where $A_{\gamma} \propto n_{i,\gamma}$ is the strength of each
independent physical scattering process (i.e., $n_{i,1}$, $n_{i,2}$,
$n_{i,3}$ denote respectively the remote and background charged
impurity density and the short-range defect density). If we now define
the net conductivity exponent $\beta(n)$ through the usual $\beta = d
\ln \sigma/ d\ln n$ definition using the total conductivity
$\sigma(n)$, then obviously $\beta(n)$ will be a complex (and opaque)
function not only of $\beta_1$, $\beta_2$, and $\beta_3$, but also of
the disorder strength $n_{i,1}$, $n_{i,2}$, $n_{i,3}$. Thus, the
extraction of a {\it universal} density exponent $\beta$ (or $\alpha$)
makes sense only when one scattering mechanism dominates ---
generically, the conductivity/mobility exponent $\beta$/$\alpha$
($=\beta -1$) is nonuniversal and depends in a complex manner both on
the individual scattering mechanism and the relative strengths of
different operational scattering mechanisms.

We consider below theoretically several different disorder potentials
which may be operational in real 2D and 3D systems, obtaining
asymptotic analytical expressions for the exponent $\alpha$ and
$\beta$ in the
process as applicable.

\subsection{Zero-range disorder}

Zero-range disorder, $V_{\bf q} \equiv V_0$ (a constant), corresponds
to pure $\delta$-function real space scatterers distributed randomly
spatially. Without any loss of generality, we can drop the
$z$-dependence of the disorder (since the electron-impurity
interaction is spatially localized), and assume the 2D electron system
to be of zero thickness in the $z$-direction interacting with the
random zero-range scatterers situated in the same plane. For 3D systems
we of course assume the scatterers to be randomly distributed three
dimensionally.

For the constant $V_{\bf q}$ model potential it is straightforward to
do the momentum integration in Eq.~(\ref{eq:eq7}) to obtain the
following results for 2D and 3D semiconductor systems and graphene:
\begin{subequations}
\begin{eqnarray}
\tau & \sim & n^0   \;\;\;\;\;\;\;\; {\rm 2D}, \\
\tau & \sim & n^{-1/3}    \;\;\; {\rm 3D}, \\
\tau & \sim & n^{-1/2}   \;\;\; {\rm graphene}.
\end{eqnarray}
\end{subequations}
We note that in graphene the mobility exponent ($\alpha \equiv
\alpha_{\mu}$) differs from the relaxation rate exponent
$\alpha_{\tau}$ by $\alpha_{\mu}=\alpha_{\tau} -1/2$. In obtaining
the density dependence of the relaxation time above we use the
standard expressions for $k_F$ and $E_F$ for parabolic 2D, 3D systems,
and graphene:
\begin{subequations}
\begin{eqnarray}
k_F & = & (2\pi n)^{1/2}; \;\;\; E_F = \hbar^2 \pi n/m: \; {\rm 2D}, \\
k_F & = & (3\pi^2 n)^{1/3}; \; E_F = \hbar^2 (3\pi^2 m)^{2/3}/2m: \;
{\rm 3D}, \\ 
k_F & = & (\pi n)^{1/2}; \;\;\;\; E_F=\hbar v_0 (\pi n)^{1/2}: \; {\rm
  graphene}.
\end{eqnarray}
\label{eq:eq11}
\end{subequations}
We use $n$ throughout to denote the relevant 2D or 3D carrier density
of the ``metal'', and $v_0$ in Eq.~(\ref{eq:eq11}c) is the constant
graphene Fermi velocity defining its linear energy-momentum
relationship, $E=\hbar v_0 k$ ($v_0 \approx 10^8$ cm/s).
We assume a spin degeneracy of 2 throughout and an additional valley
degeneracy of 2 for graphene in defining $k_F$ and $E_F$.

For zero-range $\delta$-function disorder which is equivalent to
assuming an uncorrelated white-noise disorder (often also called in
the literature, ``short-range disorder'', somewhat misleadingly), we
therefore have (we define $\alpha=\alpha_{\mu}$ always as the mobility
exponent) 
\begin{equation}
\alpha =0 \;\;({\rm 2D}); \;\;\; -1/3 \;\;({\rm 3D}); \;\;\; -1 \;\;
({\rm graphene}),
\end{equation}
and
\begin{equation}
\beta = \alpha +1 = 1 \;\;({\rm 2D}); \;\;\; 2/3 \;\;({\rm 3D});
\;\;\; 0 \;\; ({\rm graphene}).
\end{equation}
These results for 2D parabolic system and graphene are known
\cite{nine,ten} and show that the conductivity grows linearly with carrier
density in 2D systems and becomes a density independent constant in
graphene when transport is limited or dominated by zero-range
white-noise type $\delta$-function background disorder. The zero-range
transport result for the 3D systems (metals or doped semiconductors)
does not appear to be as well-known (perhaps because the density
dependence of conductivity is not of much experimental interest in 3D
systems as discussed in the Introduction of this paper) and shows
surprisingly a sub-linear $\sim n^{2/3}$ increase in $\sigma(n)$ in 3D
systems for $\delta$-correlated short-range white-noise disorder.

We now move on to the more interesting and experimentally more
relevant Coulomb disorder, and discuss a number of disorder models
pertaining to Coulomb disorder in the next two subsections.

\subsection{Unscreened (long-range) Coulomb disorder}

In most semiconductor systems (2D or 3D), the dominant background
disorder arises from quenched random charged impurities in the
environment. Thus, the bare electron-impurity interaction is invariably
the long-range $1/r$ Coulomb interaction arising from the electric
potential of the charged impurity. The charged impurity could be an
intentional dopant impurity introduced to provide the doping necessary to
create free carriers in the semiconductor or an unintentional charged
impurity invariably present even in the cleanest semiconductor. In
general, the Coulomb disorder from the random charged impurities
should be screened by the free carriers themselves so that the
effective (screened) Coulomb disorder is short-ranged (this is the
infra-red regularization necessary for handling the long range nature
of Coulomb interaction). We consider the screened Coulomb disorder in
the next three sub-sections, focusing here on the unscreened bare
Coulomb disorder for the sake of theoretical completeness.

The disorder potential $V_{\bf q}(z)$ is given by the following
equation for the unscreened Coulomb interaction:
\begin{equation}
V_{\bf q}(z) = \frac{2\pi Z e^2}{\kappa q} e^{-q|z|},
\label{eq:eq14}
\end{equation}
for 2D systems and graphene, and
\begin{equation}
V_{\bf q}(z) = \frac{4\pi Z e^2}{\kappa q^2},
\label{eq:eq15}
\end{equation}
for 3D systems.
Here $q=|{\bf q}|$ in Eqs.~(\ref{eq:eq14}) and (\ref{eq:eq15}) is the
appropriate 2D and 3D wave vector, $Z$ is the atomic number of the
charged impurity center with $Ze$ being its charge (we consider $Z=1$
throughout), and $\kappa$ is the background static lattice dielectric
constant. The coordinate $z$ in Eq.~(\ref{eq:eq14}) denotes the
spatial separation of the charged impurity from the 2D confinement
plane of the electron layer (taken to be located at $z=0$ in the
$x$-$y$ plane). We note that Eqs.~(\ref{eq:eq14}) and (\ref{eq:eq15})
are simply the Fourier transform of the $1/r$ three-dimensional
Coulomb interaction (with ${\bf r}\equiv (x,y,z)$ a 3D space vector)
in 2D and 3D systems.

For 3D systems, unscreened Coulomb disorder leads to the following
expression for the scattering rate:
\begin{eqnarray}
\tau^{-1}  =  \frac{2\pi N_i}{\hbar} & & \int \frac{d^3 k'}{(2\pi)^3} \left
    [ \frac{4\pi e^2}{\kappa |{\bf k-k'}|^2} \right ]^2 \nonumber \\
         & \times & (1-\cos \theta_{\bf kk'}) \delta(E_k-E_{k'}),
\label{eq:eq16}
\end{eqnarray}
where all wave vectors are now three dimensional and $N_i$ is the 3D
impurity density. The momentum integration in Eq.~(\ref{eq:eq16}) is
straightforward, and it is well-known \cite{eleven,twelve} that the
integral has a logarithmic divergence arising from the long-range
nature of bare Coulomb interaction (i.e., an infra-red
singularity). We get:
\begin{equation}
\tau^{-1} \sim (n \ln b)^{-1} \rightarrow \infty,
\label{eq:eq17}
\end{equation}
with $b= 4 k_F^2/q_{TF}^2 \rightarrow 0$, where $k_F = (3\pi^2
n)^{1/3}$, and $q_{TF}$, which in principle is the 3D screening wave
vector, goes to zero in the unscreened approximation, leading to a
logarithmic divergence without the screening cut-off of the long-range
Coulomb interaction. There are several ways of cutting off this
long-range logarithmic Coulomb divergence (e.g., Conwell-Weiskoff
approximation or Brooks-Herring-Dingle approximation) which has been
much discussed in the transport literature on doped semiconductors
\cite{eleven,twelve}. Since our interest in mainly focused on 2D
systems where the carrier density can be tuned continuously (in
contrast to 3D systems), we do not further discuss the implications of
Eq.~(\ref{eq:eq17}) for 3D doped semiconductors.

For 2D systems, one must distinguish among (at least) three different
kinds of Coulomb disorder: Random 2D charged impurities in the 2D
plane of the carriers [i.e., $z=0$ in Eq.~(\ref{eq:eq14})], random charged
impurities in a 2D layer parallel to the 2D carriers with a separation
$d$ (i.e. $z=d$), and 3D random charged impurity centers in the
background (where $z$
varies over a region in space). We refer to these three situations as
2D near impurities, remote impurities, and 3D impurities, respectively,
throughout this paper.

We first consider the 2D near impurity case with $z=0$ assuming unscreened
impurity-electron Coulomb interaction in the 2D plane:
\begin{equation}
V_q(z=0) = \frac{2\pi e^2}{\kappa q}.
\end{equation}
This then gives (with $n_i$ now as the 2D impurity density):
\begin{eqnarray}
\tau^{-1}  =   \frac{2\pi n_i}{\hbar} & & \int \frac{d^2k'}{(2\pi)^2} \left [
  \frac{2\pi e^2}{\kappa |{\bf k-k'}|} \right ]^2 \nonumber \\
         & \times & (1-\cos \theta_{\bf kk'}) \delta(E_k-E_{k'}).
\label{eq:eq19}
\end{eqnarray}
The 2D integral in Eq.~(\ref{eq:eq19}), in contrast to the
corresponding (divergent) 3D integral defined by Eq.~(\ref{eq:eq16}),
is convergent:
\begin{equation}
\tau^{-1} \sim k_F^{-2} \sim n^{-1}.
\label{eq:eq20}
\end{equation}
Thus $\alpha =1$ in 2D systems for the unscreened 2D impurity case
(while the corresponding 3D case is logarithmically divergent).
The exponent $\beta = \alpha + 1=2$ for the unscreened 2D Coulomb
impurity in 2D ``metallic'' systems.

Next we consider the 2D system with 2D remote impurities ($z=d \neq 0$).
The relaxation rate [Eq.~(\ref{eq:eq7})] is now given by
\begin{eqnarray}
\tau^{-1}  =   \frac{2\pi n_i}{\hbar} & & \int \frac{d^2k'}{(2\pi)^2} \left [
  \frac{2\pi e^2 e^{-|{\bf k-k'}| d}}{\kappa |{\bf k-k'}|} \right ]^2 \nonumber \\
         & \times & (1-\cos \theta_{\bf kk'}) \delta(E_k-E_{k'}).
\label{eq:eq21}
\end{eqnarray}
The integral in Eq.~(\ref{eq:eq21}) can be rewritten in a
dimensionless form
\begin{equation}
\tau^{-1} \sim k_F^{-2} \int_0^1 dx\frac{e^{-2x d_0}}{\sqrt{1-x^2}},
\label{eq:eq22}
\end{equation}
where $d_0 = 2k_F d$ is dimensionless. [We note that putting
$d_0=0$ for $d=0$ gives us the 2D near impurity results of
  Eq.~(\ref{eq:eq20}).] 
The asymptotic carrier density dependence ($k_F \propto \sqrt{n}$)
implied by Eq.~(\ref{eq:eq22}) depends sensitively on whether
$k_Fd$ [$\equiv d_0$ in Eq.~(\ref{eq:eq22})] is small ($k_F d \ll
1$) or large ($k_F d \gg 1$).
For $k_Fd \ll 1$, we get from Eq.~(\ref{eq:eq22})
\begin{equation}
\tau^{-1} \sim k_F^{-2} \sim n^{-1}
\end{equation}
and for $k_Fd \gg 1$, we get from Eq.~(\ref{eq:eq22})
\begin{equation}
\tau^{-1}\sim k_F^{-3} \sim n^{-3/2}.
\end{equation}
Thus, $\alpha =1$ ($3/2$) for $k_F d \ll 1$ ($\gg 1$) and therefore
$\beta = 2$ ($5/2$) for $k_Fd \ll 1$ ($\gg 1$) for 2D carriers in the
presence of remote Coulomb scatterers.

Finally, we consider the 2D carrier system with a 3D random
distribution of charged impurity centers. The integral for the 2D
relaxation rate (Eq.~(\ref{eq:eq7})) now becomes with $q_{TF}
\rightarrow 0$
\begin{equation}
\tau^{-1}\sim k_F^{-3} \ln\left ( \frac{q_{TF}}{2k_F} \right ) \sim
n^{-3/2} \ln \left ( \frac{q_{TF}}{2k_F} \right ),
\end{equation}
which has the same logarithmic divergence for the unscreened Coulomb
disorder as the corresponding 3D case considered above in
Eq.~(\ref{eq:eq17}), but with a different 2D density exponent ($\sim
n^{-3/2}$) 
from the corresponding 3D density exponent ($\sim n^{-1/3}$) in
Eq.~(\ref{eq:eq17}). Thus, 2D carrier systems with unscreened 3D Coulomb
disorder would have logarithmically divergent resistivity
necessitating a length cut-off on the long range part of the bare
Coulomb potential similar to the well-known situation for long-range
bare Coulomb disorder in 3D systems (e.g. doped semiconductors).

It is important to emphasize our interesting theoretical finding that,
although 3D unscreened Coulomb disorder leads to logarithmic
long-range divergence in the resistivity of both 2D and 3D systems
independent of the carrier density (i.e. $\tau^{-1}$ diverges
logarithmically), the corresponding situation for 2D carrier systems
with 2D Coulomb disorder has no divergence and does not require any
cut-off dependent infra-red regularization. It turns out that 2D 
metals with unscreened Coulomb disorder arising from random 2D charged
impurities is perfectly well-defined within the Boltzmann transport
theory. Of course, the unscreened Coulomb disorder model is not
realistic and the calculated conductivity may not agree with the
experimental data, but theoretically it is perfectly well-defined with
the density exponents $\alpha$ ($\beta$) being 1 (3/2) and 2 (5/2),
respectively depending on whether $k_Fd \ll 1$ or $\gg 1$ where $d$ is
the location of the impurities with respect to the 2D carrier layer.

One may wonder about the fundamental reason underlying the necessity for
infra-red regularization for the 3D unscreened case and {\it not} for
the 2D unscreened case. This arises from the 3D bare Coulomb potential
($\sim 1/q^2$) being much more singular in the long wavelength
$q\rightarrow 0$ limit than the corresponding 2D Coulomb potential
($\sim 1/q$). It turns out that this difference between 2D and 3D is
sufficient to make the 3D Coulomb disorder case infra-red divergent
whereas the 2D case being non-divergent without infra-red
regularization.

We do not discuss here the unscreened Coulomb disorder case for
graphene since for graphene the density scaling of the conductivity is
the same
for both unscreened and screened Coulomb disorder (and we consider
screened Coulomb disorder in the next subsection) by virtue of
graphene screening wave vector $q_{TF}$ ($\propto k_F$) being
proportional to the Fermi wave vector leading to both screened and
unscreened Coulomb disorder having the same carrier density dependence
in the conductivity.

\subsection{Screened Coulomb Disorder}

This is the most realistic (as well as reasonably computationally
tractable) model for calculating the resistivity due to Coulomb
scattering within the Boltzmann transport theory. The basic idea is to
use the appropriately screened Coulomb potential in Eq.~(\ref{eq:eq7}) for
calculating the relaxation rate. We use the conceptually simplest
(static) random phase approximation (RPA) for carrier screening of the
long-range electron-impurity Coulomb interaction. This means that the
screening functions in 2D and 3D parabolic systems are 
the static dielectric functions first calculated by Stern
\cite{thirteen} and Linhard \cite{fourteen}, respectively. For
graphene, we use the dielectric screening function first calculated in
Ref.~[\onlinecite{fifteen}].

The relevant screened Coulomb disorder potential is given by:
\begin{equation}
V_{q} \equiv \frac{v_q}{\epsilon(q)},
\end{equation}
where $v_q$ is the long-range Coulomb potential and $\epsilon(q)$ is
the appropriate static RPA dielectric function. In general, the
screened Coulomb potential can be rewritten as
\begin{equation}
V_q = \frac{2\pi e^2}{\kappa (q + q_{TF})},
\end{equation}
for 2D systems and graphene, and
\begin{equation}
V_q = \frac{4\pi e^2}{\kappa (q^2 + q_{TF}^2)},
\end{equation}
for 3D systems. The screening wave vector $q_{TF}$, sometimes referred
to as the Thomas-Fermi wave vector is given by the following
expression (obtained in a straightforward manner from the
corresponding static polarizability function or the dielectric
function in Refs. \onlinecite{thirteen,fourteen,fifteen}):
\begin{subequations}
\begin{eqnarray}
q_{TF} & = & \frac{2 m e^2}{\kappa \hbar^2} \;\;\;\;\;\;\; {\rm 2D}, \\
q_{TF} & = & \left ( \frac{4me^2 k_F}{\pi \kappa \hbar^2} \right )^{1/2} \;
{\rm 3D}, \\ 
q_{TF} & = & \frac{4 e^2 k_F}{\kappa \hbar v_0} \;\;\;\;\;\; {\rm graphene}.
\end{eqnarray}
\label{eq:eq28}
\end{subequations}
We have used a valley degeneracy of 1 (2) for 2D/3D (graphene) systems
in Eqs.~(\ref{eq:eq28}), and chosen a spin degeneracy of 2. We note the
(well-known) results that in the 2D parabolic electron system the
screening wave vector is a constant, whereas in graphene (3D parabolic
system) it is proportional to $k_F$ ($\sqrt{k_F}$).

It may be worthwhile to discuss the dimensionless screening strength
characterized by the parameter $q_s=q_{TF}/2k_F$, which is given by
\begin{subequations}
\begin{eqnarray}
q_{s} & = & \frac{2 m e^2}{\kappa \hbar^2k_F} \sim n^{-1/2} \;\;\;\;\; {\rm 2D}, \\
q_{s} & = & \left ( \frac{4me^2}{\pi \kappa \hbar^2 k_F} \right )^{1/2} \sim n^{-1/6}
\;\; {\rm 3D}, \\ 
q_{s} & = & \frac{4 e^2}{\kappa \hbar v_0} \sim n^0 \;\;\;\;\; {\rm graphene}.
\end{eqnarray}
\label{eq:eq31}
\end{subequations}
We note the curious (albeit well-established) result that the
dimensionless screening strength increases (very slowly in 3D $\sim
n^{-1/6}$) with decreasing density ($\sim n^{-1/2}$ in 2D) except in
graphene where $q_{TF} \propto k_F$ leading to a density-independent
$q_s$. Thus, the strong (weak) screening limit with $q_s =q_{TF}/2k_F
\gg 1$ ($\ll 1$) is reached at low (high) carrier density in 2D and 3D
metallic systems. This peculiar density-dependence of screening has
qualitative repercussion for the transport exponents $\alpha$ and
$\beta$ as a function of density.

We now rewrite Eq.~(\ref{eq:eq7}) for the 2D relaxation rate $\tau^{-1}$ in
terms of screened Coulomb disorder obtaining;
\begin{equation}
\tau^{-1} = \left (\frac{n_i m}{\pi \hbar^3 k_F^2} \right ) \left
(\frac{2\pi e^2}{\kappa} \right )^2 I_{22}(q_s,d_0),
\label{eq:eq34}
\end{equation}
with 
\begin{equation}
I_{22}(q_s,d_0) = \int_0^1 \frac{e^{-2 x d_0} x^2 dx}{(x+q_s)^2
  \sqrt{1-x^2}}, 
\label{eq:eq35}
\end{equation}
for 2D carriers and 2D impurities (with impurity density $n_i$ per unit
area) with $q_s = q_{TF}/2k_F$ and $d_0 = 2k_F d$ (with $z=d$ defining
the impurity locations).
\begin{equation}
\tau^{-1} = \left ( \frac{N_i m}{8\pi \hbar^3 k_F^3} \right )
\left(\frac{2\pi e^2}{\kappa} \right )^2 I_{23}(q_s),
\end{equation}
with
\begin{equation}
I_{23}(q_s) = \int_0^1 \frac{dx}{(q_s + \sqrt{1-x^2})^2},
\end{equation}
for 2D carriers and 3D impurities (with impurity density $N_i$ per unit
volume). We note that putting $q_s=0$ (i.e. no screening) immediately
produces the density scaling exponents obtained in the last
subsection.

For 3D carriers with (obviously) 3D random charged impurity
distribution, we get
\begin{equation}
\tau^{-1} = \left (\frac{N_i m}{8\pi \hbar^3 k_F^3} \right )
I_{33}(q_s),
\end{equation}
where
\begin{equation}
I_{33}(q_s) = \int_0^1 dx \frac{1-x}{(1-x+2q_s^2)^2}.
\end{equation}
Again, putting $q_s=0$ in the 3D result above produces the
logarithmically divergent relaxation rate discussed in Sec. IIIB for
the 3D unscreened Coulomb disorder.

Finally, for graphene we write down the relaxation rates for
scattering by screened Coulomb disorder arising 2D and 3D charged
impurity distributions respectively by following the standard
references \cite{nine,sixteen}:
\begin{equation}
\tau^{-1} =\left( \frac{n_i}{\pi \hbar^2 v_0 k_F} \right ) \left(
\frac{2\pi e^2}{\kappa} \right )^2 I_{G2}(q_s,d_0),
\end{equation}
where
\begin{equation}
I_{G2} = \int_0^1 dx \frac{x^2 \sqrt{1-x^2}}{(x+q_s)^2}e^{-2xd_0},
\label{eq:eq41}
\end{equation}
for 2D charged impurities located a distance $d$ (with $d_0 = 2k_F
d$) from the graphene plane, and for 3D disorder:
\begin{equation}
\tau^{-1} = \left ( \frac{N_i}{\pi \hbar^2 v_0 k_F^2} \right ) \left (
\frac{2\pi e^2}{\kappa} \right )^2 I_{G3}(q_s),
\end{equation}
with
\begin{equation}
I_{G3} = \int_0^1 dx \frac{x\sqrt{1-x^2}}{(x+q_s)^2},
\label{eq:eq43}
\end{equation}
and $N_i$ being the 3D impurity density.
In the 3D Coulomb disorder case for graphene, the random charged
impurities are distributed in the graphene substrate with a uniform
random 3D distribution with a 3D impurity density of $N_i$.

For our numerical calculations of transport properties (to be
presented in the next section) which would focus entirely on 2D
systems with 2D impurities (both near and remote), we will include the
realistic width of the quantum well through a subband form-factor
modifying the Coulomb matrix element arising from the finite thickness
of the quantum well in the $z$-direction. This is a nonessential
complication (making our numerical results compatible with and
comparable with the experimental low-temperature transport data in
GaAs quantum wells) which does not affect our theoretical conclusions
about the carrier density scaling of the 2D transport properties since
the quasi-2D quantum well form factor is independent of the carrier
density in the leading order.

Below we obtain the asymptotic density exponents (based on the results
given above) $\alpha$ and $\beta$ for screened Coulomb disorder in the
strong ($q_s \gg 1$) and weak ($q_s \ll 1$) screening situations
considering both near and remote 2D impurities and 3D impurities.

\begin{table*}
\begin{tabular}{|m{5cm}||m{2.4cm}|c|c|c|}
\hline
\multicolumn{2}{|m{4cm}|}{ }    & 2D & 3D  & Graphene  \bigstrut \\  \hline \hline
\multirow{3}{*}{2D Coulomb disorder with near}
& strong screening ($q_s \gg 1$) & 0  & N/A & 0           \bigstrut \\  \cline{2-5}
{impurities ($2k_F d \ll 1$) }    & weak screening ($q_s \ll 1$)  & 1   & N/A   & 0           \bigstrut \\  \cline{2-5}
  & unscreened ($q_s=0$)     &  1  &  N/A  &  0         \bigstrut \\  \hline
\multirow{3}{*}{2D Coulomb disorder with }  & strong
screening ($q_s \gg 1$) & 3/2  & N/A & 1/2           \bigstrut \\  \cline{2-5}
{remote impurities ($2k_F d  \gg 1$)}  & weak screening ($q_s \ll 1$)  & 3/2   & N/A   &  1/2         \bigstrut \\  \cline{2-5}
  & unscreened ($q_s=0$)     &  3/2  &  N/A  &     1/2      \bigstrut \\  \hline
\multirow{3}{*}{3D Coulomb disorder with 3D} 
& strong screening & 1/2  & 1/3 & 1/2         \bigstrut \\  \cline{2-5}
& weak screening   &  3/2  & 1   &  1/2         \bigstrut \\  \cline{2-5}
{impurity distribution}      & unscreened      & log-divergent   &  log-divergent  & log-divergent          \bigstrut \\  \hline
zero-range disorder with $\delta$-function impurities & concept of
screening inapplicable here & 0 & $-1/3$ & $-1$ \\ \hline
\end{tabular}
\caption{Asymptotic values of $\alpha$. By considering the
Coulomb disorder arising from random charged impurities in the
environment the asymptotic density scaling exponent
($\alpha$) for the carrier mobility ($\mu \sim n^{\alpha}$) in various
limiting situations and for various types of disorder is given. The
corresponding conductivity exponent ($\beta$) with $\sigma \sim
n^{\beta}$ is given by $\beta = \alpha + 1$. 
\label{tab:table1}}
\hspace{0.1\hsize}
\end{table*}

\subsection{Strong-screening ($q_s \gg 1$) and weak-screening ($q_s
  \ll 1$) limits}

It is straightforward to carry out the asymptotic expansions of the
various integrals in
Eqs.~(\ref{eq:eq34})--(\ref{eq:eq43}) to obtain the strong-screening ($q_s \gg1$)
and the weak-screening ($q_s \ll 1$) limiting behaviors of the
relaxation rate $\tau^{-1}$ for the different systems under
consideration. Remembering that $\tau^{-1} \sim n^{-\alpha}$ and $\beta
= \alpha + 1$ ($\alpha + 1/2$ for graphene) we get the following
results by taking $q_s \gg 1$ and $q_s \ll 1$ limits of
Eqs.~(\ref{eq:eq34})--(\ref{eq:eq43}) in various situations. \\

\noindent 
(i) 2D carriers with 2D impurities:\\
For $2k_Fd \ll 1$ (i.e. near impurities) \\
$\alpha = 0$ ($\beta = 1$) for strong screening ($q_{TF} \gg 2k_F$) \\
$\alpha = 1$ ($\beta =2$) for weak screening ($q_{TF} \ll 2k_F$). \\
For $2k_F d \gg 1$ (i.e. remote impurities) \\
$\alpha = 3/2$ ($\beta = 5/2$) for both weak ($q_{TF} \ll 2k_F$) and
strong ($q_{TF} \gg 2k_F$) screening. \\

\noindent
(ii) 2D carriers with 3D impurities: \\
$\alpha = 1/2$ ($\beta = 3/2$) for strong screening ($q_{TF} \gg
2k_F$) \\
$\alpha = 3/2$ ($\beta = 5/2$) for weak screening ($q_{TF} \ll 2k_F$).
\\

\noindent
(iii) 3D carriers with 3D impurities: \\
$\alpha=1/3$ ($\beta = 4/3$) for strong screening ($q_{TF} \gg 2k_F$)
\\
$\alpha=1$ ($\beta=2$) for weak screening ($q_{TF} \ll 2k_F$). \\

\noindent
(iv) Graphene with 2D impurities (remembering $\alpha =
\alpha_{\tau}-1/2 = \alpha_{\mu}$): \\
For $2k_F d \ll 1$ (i.e. near impurities)
$\alpha_{\tau} = 1/2$; $\alpha = 0$ ($\beta = 1$) for both strong and
weak screening. \\
For $2k_Fd \gg 1$ (i.e., remote impurities)
$\alpha_{\tau}=1$; $\alpha = 1/2$ ($\beta = 3/2$) for both strong and
weak screening. \\
We note that for graphene $\alpha = \alpha_{\mu} = \alpha_{\tau} -
1/2$ (and $\beta = \alpha + 1$) by virtue of its linear
dispersion. Also, for graphene strongly screened, weakly screened and
unscreened Coulomb disorder manifest the same density exponent in the
conductivity and the mobility since $q_{TF} \propto k_F$, and thus
$q_s$ is density independent. (Of course, the actual numerical values
of the conductivity and the mobility are very different in the three
approximations except for having the same power law dependence on the
carrier density depending only on whether the random 2D charged
impurities are near or far.) \\

\noindent
(v) Graphene with screened 3D impurities:
$\alpha_{\tau}=1$; $\alpha = 1/2$ ($\beta = 3/2$) for both strong and
weak screening. \\
We note that graphene with unscreened 3D Coulomb disorder ($q_s=0$)
has the same logarithmic divergence in the relaxation rate $\tau^{-1}$
(and hence in the resistivity) for all carrier densities as in the
corresponding 2D and 3D parabolic electron systems for unscreened 3D
disorder. The unscreened 3D Coulomb disorder is thus unphysical,
always necessitating an infra-red regularization as was already
realized in the 1950s in the context of 3D doped semiconductor
transport \cite{eleven,twelve}.


In Table I we summarize our asymptotic analytic findings for the
density scaling of conductivity and mobility for 2D, 3D parabolic
systems and 2D graphene in various limiting situations involving
Coulomb disorder arising from random charged impurities in the
environment.

In Table I above we provide the asymptotic density scaling exponent
($\alpha$) for the carrier mobility ($\mu \sim n^{\alpha}$) in various
limiting situations and for various types of disorder. The
corresponding conductivity exponent ($\beta$) with $\sigma \sim
n^{\beta}$ is given by $\beta = \alpha + 1$.

\section{results}

In this section we present our numerical results for the $T=0$ density
dependent transport properties of 2D systems within the Boltzmann
transport theory in order to obtain the full density dependence of the
exponent $\alpha(n)$ for mobility and equivalently the exponent
$\beta(n)$ for conductivity, obtaining in the process the density regimes
where the analytical asymptotic exponents obtaining in the last
section apply. The reason we focus on the 2D carrier system is that
it is the most convenient system for the experimental investigation of
the density dependence of transport properties. In 3D semiconductor
systems, the carrier density cannot be continuously tuned as it can be
in 2D systems.

We first note that the strong (weak) screening condition implies low
(high) values of carrier density in the system. Using
Eq.~(\ref{eq:eq11}) and
Eq.~(\ref{eq:eq28}), we find: 
$q_s = q_{TF}/2k_F \gg 1$ implies $2m e^2/ \kappa \hbar^2 \gg  2 (2
\pi n)^{1/2}$: 2D, $(4 me^2/\pi \kappa \hbar^2)^{1/2} (3 \pi^2
n)^{1/6} \gg 2 (3\pi^2 n)^{1/3}$: 3D, $(4e^2 k_F/\kappa \hbar v_0) \gg
2 k_F$: graphene, i.e.,
\begin{subequations}
\begin{eqnarray}
\frac{1}{8\pi}\left ( \frac{2 m e^2}{\kappa \hbar^2} \right )^2 & \gg & n \;\;\;
     {\rm 2D}, \\ 
\frac{1}{2\pi^2}\left ( \frac{4me^2}{\pi \kappa \hbar^2} \right )^3
& \gg & n \;\;\; {\rm 3D}, \\ 
\left (\frac{2 e^2}{\kappa \hbar} \right ) & \gg & v_0  \;\;\; {\rm graphene}.
\end{eqnarray}
\label{eq:eq44}
\end{subequations}
Eq.~(\ref{eq:eq44}) define the low-density regime where the strong
screening condition would be satisfied except for graphene which has
$q_s$ independent of carrier density since $q_{TF} \propto k_F$. From
Eq.~(\ref{eq:eq44}) we conclude that strong (weak) screening situation
(within RPA) for $T=0$ transport properties would be satisfied under the
following conditions for the different systems under consideration:
\begin{subequations}
\begin{eqnarray}
n & \ll & \; (\gg) \left ( \frac{m}{m_e \kappa} \right )^2 \times 1.14
\times 10^{16} cm^{-2} \;\;
     {\rm 2D}, \\ 
n & \ll & \; (\gg) \left ( \frac{m}{m_e \kappa} \right )^3 \times 7.4
\times 10^{21} cm^{-2} \;\; {\rm 3D}, \\ 
\kappa & \ll & \; (\gg) \;\; 4.4
\;\;\;\;\; {\rm graphene}.
\end{eqnarray}
\label{eq:eq47}
\end{subequations}
Using the band effective mass for GaAs electrons ($m=0.07 m_e$) and
holes ($m=0.4m_e$), we get (using $\kappa=13$ for GaAs-AlGaAs quantum
wells)
\begin{subequations}
\begin{eqnarray}
n &\ll& \; (\gg) \; 3.3 \times 10^{11} cm^{-2} \;\;
     {\rm for \; 2D \; n-GaAs}, \\ 
n &\ll& \; (\gg) \; 9.5 \times 10^{13} cm^{-2} \;\; {\rm for \; 2D \; p-GaAs}, 
\end{eqnarray}
\label{eq:eq471}
\end{subequations}
and
\begin{subequations}
\begin{eqnarray}
n &\ll& \; (\gg) \; 1.5 \times 10^{15} cm^{-3} \;\;
     {\rm for \; 3D \; n-GaAs}, \\ 
n &\ll& \; (\gg) \; 2.8 \times 10^{17} cm^{-3} \;\; {\rm for \; 3D \; p-GaAs}.
\end{eqnarray}
\label{eq:eq481}
\end{subequations}
We note (again) that in graphene (as Table I indicates) the exponents
$\alpha$, $\beta$ do not depend on weak/strong screening or on the
carrier density. The density range for GaAs-based 2D systems, which we
would consider numerically in this section, is the experimentally
relevant $10^9-10^{12}$ cm$^{-2}$ density range in high-mobility
GaAs-Al$_x$Ga$_{1-x}$As 2D quantum well structures, and thus for 2D
electron systems (2D n-GaAs) the crossover from the low-density
strong-screening to high-density weak-screening behavior may be
experimentally relevant. On the other hand, for 2D p-GaAs hole quantum
wells, the crossover density ($\sim 10^{14}$ cm$^{-2}$) is too high to
be relevant experimentally, and thus the laboratory 2D hole systems
are likely to be always
in the strongly screened situation.

\begin{figure}[t]
\includegraphics[width=1.0\columnwidth]{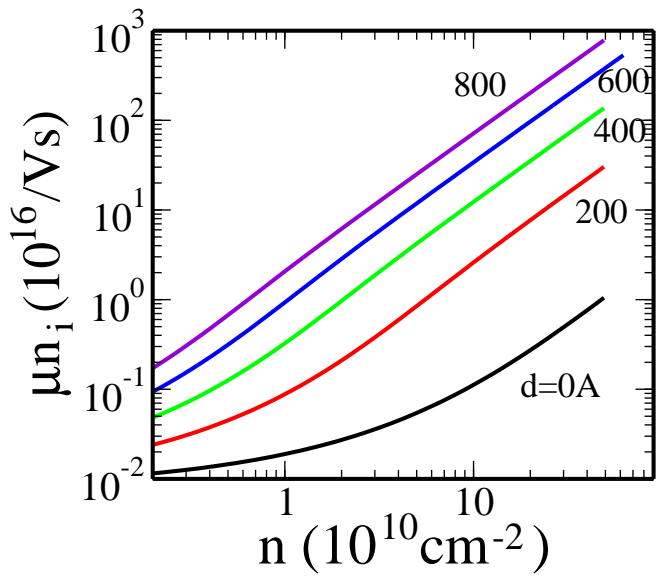}
\caption{(color online) 
Calculated mobility, $\mu n_i$, as a function of density $n$ of n-GaAs
quantum well with a well width $a=200$ \AA \; for various $d=$0,
200, 400, 600, 800 \AA (from bottom to top).  
Here $n_i$ is the 2D random charged
impurity density located a
distance $d$ away from the quantum well.
The mobility is calculated at $T=0$ K.
\label{fig:one}
}
\end{figure}
\begin{figure}[t]
\includegraphics[width=.90\columnwidth]{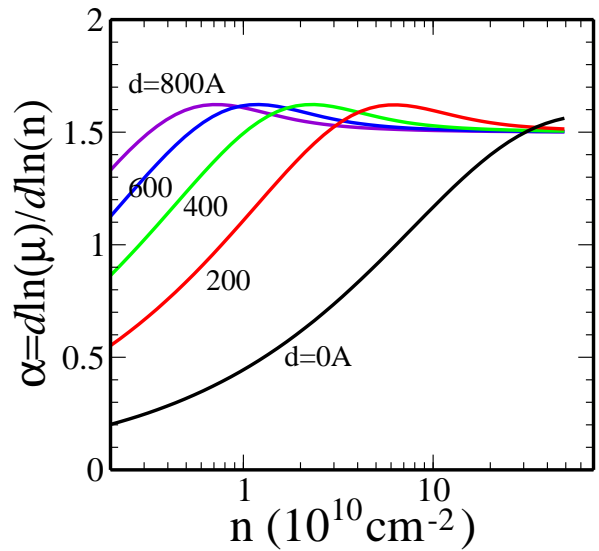}
\caption{(color online) 
Calculated mobility exponent $\alpha$ in $\mu \propto n^{\alpha}$
(i.e., $\alpha = d \ln \mu /d\ln n$) for the corresponding $\mu(n)$ in
Fig.~\ref{fig:one}. 
\label{fig:two}
}
\end{figure}

The 2D transport exponents $\alpha$ and $\beta$ depend on an
additional dimensionless parameter (i.e. in addition to $q_s =
q_{TF}/2k_F$) $d_0=2k_F d$, which depends both on the carrier density
through $k_F \sim \sqrt{n}$ and on the separation ($d$) of the random
charged impurities from the 2D system. This dependence on $k_Fd$, the
dimensionless separation of the impurities from the carriers in the
$z$ direction (which has no analog in the corresponding 3D disordered
case), is experimentally always relevant in high-mobility 2D
semiconductor structures for three reasons: (1) in high-mobility
modulation-doped 2D quantum well structures, scattering by remote
dopants (which are introduced intentionally at a known distance $d$
from the quantum well) is always present; (2) in principle, it is
always possible to place ionized impurities at a set distance $d$ from
the 2D quantum well using the computer-controlled MBE growth technique
which is used to produce high-quality semiconductor quantum wells to
begin with; (3) in realistic experimental samples with a finite
quasi-2D thickness, $d$ is always finite.

In general, the experimental 2D mobility/conductivity could be limited
by various types of Coulomb disorder \cite{seventeen}: near/far 2D
Coulomb disorder ($2k_F d \ll 1$/ $\gg 1$): 3D Coulomb disorder in the
background. According to Table I, the asymptotic density dependence in
each case should be different. In Figs.~\ref{fig:one} and
\ref{fig:two} we show our directly numerically calculated n-2D GaAs
mobility for quantum well electrons assuming 2D random charged
impurity scattering from quenched point scattering centers located a
distance $d$ away from the quantum well (including the $d=0$
case). The numerically calculated mobility exponent $\alpha(n) = d \ln
\mu/d\ln n$ is shown in Fig.~\ref{fig:two} for the corresponding
$\mu(n)$ shown in Fig.~\ref{fig:one} -- we note that $n \propto
k_F^2$, and therefore the parameter $k_F d \sim \sqrt{n}$ for a fixed
value of $d$. The results shown in Figs.~\ref{fig:one} and
\ref{fig:two} correspond to the zero-temperature case, but are
indistinguishable from the corresponding low-temperature results for
$T=300mK$ (which we have verified explicitly). We note that the
numerical results presented in Figs.~\ref{fig:one} and \ref{fig:two}
are realistic (as are all other numerical results shown in this paper) in the
sense that they are obtained from the full numerical integration of
the Boltzmann theory expression for the relaxation rate [as given in
Eq.~(\ref{eq:eq7}) in section III] with the additional sophistication of
including the finite thickness of the quantum well through the finite
well-thickness form-factors $f_i(q)$ and $f(q)$ which modify the
Coulomb disorder 
matrix element (i.e. $|V|^2 \rightarrow |V|^2 f_i(q)$) and  $q_{TF}
\rightarrow q_{TF} f(q)$, respectively, and they are given by
\begin{eqnarray}
f_i(q) & = & \frac{4}{qa} \frac{2\pi^2(1-e^{-qa/2}) + (qa)^2}{4\pi^2 +
  (qa)^2}, \nonumber \\
f(q) & = & \frac{3(qa)+8\pi^2/(qa)}{(qa)^2+4\pi^2} -
\frac{32\pi^4[1-\exp(-qa)]}{(qa)^2[(qa)^2+4\pi^2]^2}.
\end{eqnarray}
where $a$ refers to the quantum well width taken to be 200 \AA \; for
the results in Figs.~\ref{fig:one} and \ref{fig:two}. We note that the
form-factor $f_i(q)$ simply reduces the Coulomb impurity potential from its
$q^{-1}$ behavior by a $q$-dependent (but density-independent)
function determined by the thickness $a$ of the GaAs quantum
well. The form-factor $f(q)$ reduces the 2D screening through the
modification of the electron-electron interaction due to the finite
thickness of the quantum well.
Since the quantum well form factors do not depend on the carrier
density in the leading order, this quasi-2D form-factor effect does
not in any way modify the asymptotic exponents $\alpha$ and $\beta$
given in Table I (and theoretically defined in the lase section), but
the form factors do modify the actual calculated values of the
mobility/conductivity/resistivity, making them more realistic, and
therefore any comparison with experimental transport data necessitates
the inclusion of the quasi-2D form factor effect. All our numerical
results for 2D electrons and holes in GaAs quantum wells presented in
this paper include the realistic quantum well form factors in the
theory taking into account the finite well width effect both in the
electron-impurity interaction and in the electron screening [with
$q_{TF}$ being modified to $q_{TF} f(q)$].

\begin{figure}[t]
\includegraphics[width=1.0\columnwidth]{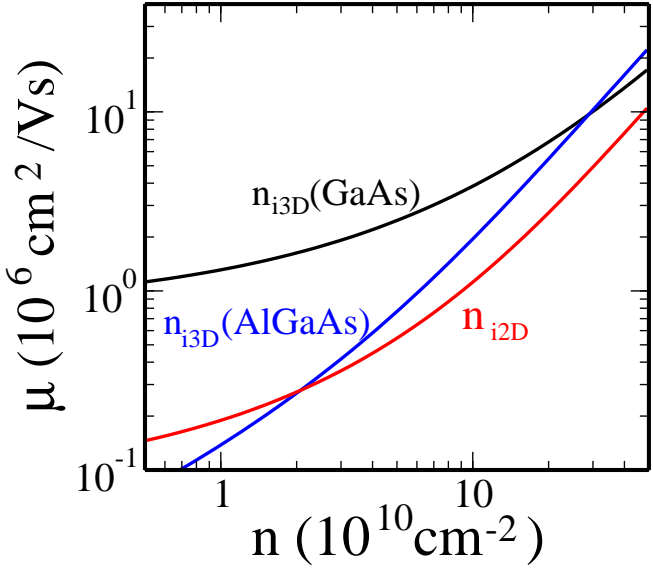}
\caption{(color online) 
Calculated mobility of n-GaAs quantum well ($a=200$\AA) with unintentional
charged impurities. Here the 2D
impurities ($n_{i2D}=10^9$ cm$^{-2}$) are at the GaAs-AlGaAs interface
whereas there are two different types of 3D Coulomb disorder: inside
the GaAs well [$n_{i3D}=10^{14}$ cm$^{-3}$ (GaAs)] and inside the barrier AlGaAs
[$n_{i3D}=10^{15}$ cm$^{-3}$(AlGaAs)]. 
\label{fig:three}
}
\end{figure}
\begin{figure}[t]
\includegraphics[width=.9\columnwidth]{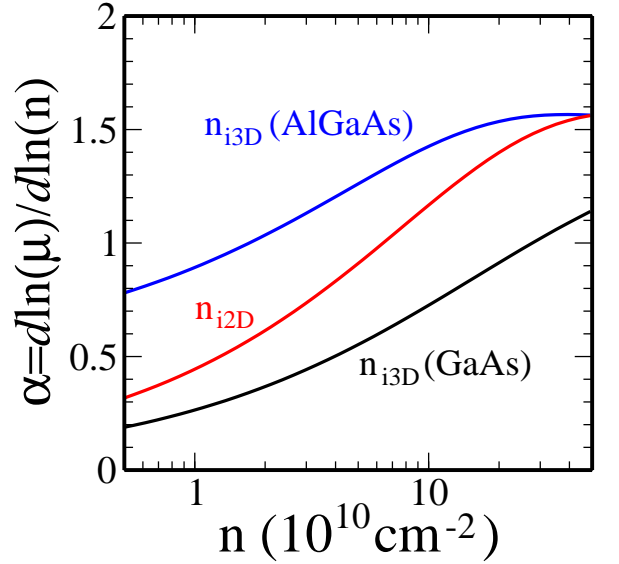}
\caption{(color online) 
Calculated mobility exponent $\alpha$ in $\mu \propto n^{\alpha}$
(i.e., $\alpha = d \ln \mu /d\ln n$) for the corresponding $\mu(n)$ in
Fig.~\ref{fig:three}. 
\label{fig:four}
}
\end{figure}

In Fig.~\ref{fig:one} we show (we actually show the calculated $\mu
n_i$ since $\mu \sim 1/n_i$) our calculated mobility $\mu$ for 2D n-GaAs
as a function of carrier density $n$ for different impurity locations
($d$) whereas in Fig.~\ref{fig:two} we show the corresponding mobility
exponent $\alpha(n)=d\ln \mu/d\ln n$. The asymptotic high-density
results, $\alpha \rightarrow 3/2$ for $n \rightarrow \infty$ (i.e., $q_{TF}
\ll 2k_F$) and $k_F d \gg 1$ as given in Table I, is clearly obeyed in
all cases with $\alpha \approx 3/2$ for larger $d$ and $n$ values
satisfying $2k_F d \gg 1$. For 2D GaAs systems:
\begin{equation}
2 k_F d \approx 5 \tilde{d} \sqrt{\tilde{n}},
\end{equation}
where $\tilde{d}$ is measured in 1000 \AA \; units and $\tilde{n}$ in
units of $10^{10}$ cm$^{-2}$.
Thus even for $d=100$\AA, $2k_F d \agt 1$ already for $n \agt 2 \times
10^{10}$ cm$^{-2}$, and thus the $2k_F d \gg 1$ condition is quickly
reached at higher carrier density in all 2D GaAs systems for any type
of relevant background 2D Coulomb disorder (since typically, $n \approx
10^9$ cm$^{-2}$ is a lower limit for the achievable carrier density in
2D semiconductor systems). One may wonder if the $d=0$ situation
corresponds to the $2k_F d \equiv 0$ situation considered in Table
I. This is certainly true for the strict 2D limit (i.e. $a=0$ limit of
the quantum well). But for any finite value of $a$, the $d=0$ impurity
location only refers to the distance of the 2D charged impurities from
the GaAs-AlGaAs interface, and thus the average impurity separation
from the electrons is always finite except in the $a \rightarrow 0$
limit. Thus even the $d=0$ case in our theoretical calculation has an
effective finite value of $d_0 = 2k_F d$ because of the finite layer
thickness effect (e.g., $d \approx a/2$ effectively in the $d
\rightarrow 0$ limit).

The most important conclusions from Figs.~\ref{fig:one} and
\ref{fig:two} are: (i) The $2k_Fd \gg 1$ condition dominates the
mobility exponent except for rather low mobility samples with very
small values of $d$; (ii) even for the nominal $d=0$ case in Fig. 1
(where in the strict 2D case, $\alpha \alt 1$  always) $\alpha$ eventually
approaches the asymptotic $\alpha \rightarrow 3/2$ value for $n \gg
10^{11}$ cm$^{-2}$ because of the finite layer thickness effect (i.e.,
finite $a$); (iv) for scattering purely by very remote dopants ($d
\agt 500$ \AA), the mobility exponent $\alpha > 1$ always because $2k_F
d \gg 1$ condition is always satisfied; (v) the low density limit
($2k_F d \ll 1$, $q_s \gg 1$), where $\alpha \rightarrow 0$ according
to Table I, would be achieved in 2D n-GaAs systems only for $n \ll
10^9$ cm$^{-2}$, and in all realistic situations, $\alpha > 0.5$
always as long as transport is dominated by 2D Coulomb disorder.

In Fig.~\ref{fig:three} and \ref{fig:four} we show (again for $a=200$
\AA) our calculated mobility in the presence of both 2D and 3D
(unintentional background) disorder neglecting remote scattering
effects (assuming the intentional remote dopants to be too far, $d >
1000$ \AA, for them to have any quantitative effects, as would apply
to gated undoped HIGFET structures or to extreme high-mobility
modulation doped structures where the intentional dopants are placed
very far away). In Figs.~\ref{fig:three} and \ref{fig:four}, 
the 2D
impurities ($n_{i2D}$) are put right at the GaAs-AlGaAs interface
whereas there are two different types of 3D Coulomb disorder: inside
the GaAs well [$n_{i3D}$ (GaAs)] and inside the barrier AlGaAs
[$n_{i3D}$ (AlGaAs)]. Again the corresponding critical exponents
increase with carrier density, approaching $\alpha =3/2$ for high
density consistent with the analytic theory. When
the dominant background disorder is that arising from $n_{i3D}$(GaAs),
i.e., unintentional background impurities in the well itself,
typically $\alpha \approx 0.5-0.8$, which is in between weak and
strong screening situation.

\begin{figure}[t]
\includegraphics[width=1.0\columnwidth]{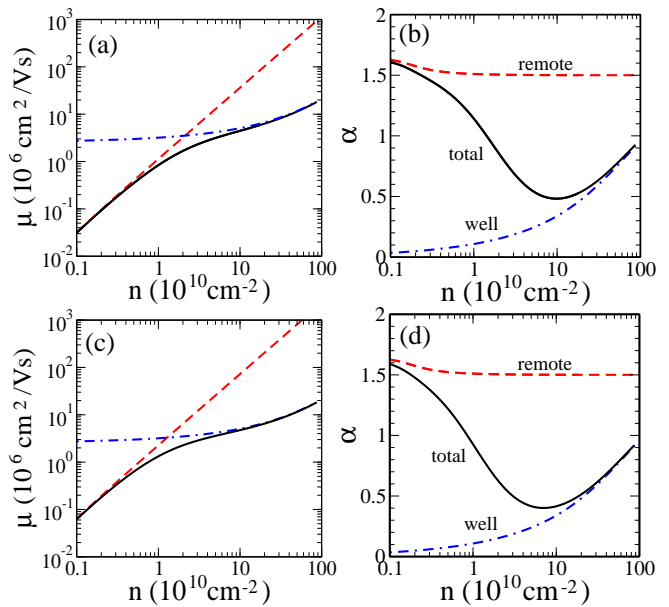}
\caption{(color online) 
(a) Hole mobility as a function of hole density ($n$) of
p-GaAs quantum well with a width $a=200$ \AA \; and the hole mass of
$m=0.3$. Here the background 
unintentional charged impurities with a density
$n_{i3D}=3 \times 10^{13} cm^{-3}$ are located inside the quantum well 
and 2D remote charged impurities with a density 
$n_i=8 \times 10^{11} cm^{-2}$ are located at $d= 150$ \AA \; from the
interface. In this figure
the black solid curve indicates the total mobility and blue dot-dashed
(red dashed) curve indicates the  
mobility limited by only background scattering (remote charged scattering). 
(b) The exponents ($\alpha$) for the corresponding mobilities in (a).
(c) Hole mobility with the same parameters of Fig.~\ref{fig:five}(a) except the
remote charged impurity density  
$n_i=4 \times 10^{11} cm^{-2}$.
(d) The exponents ($\alpha$) of corresponding mobilities of (c).
\label{fig:five}
}
\end{figure}

\begin{figure}[t]
\includegraphics[width=1.0\columnwidth]{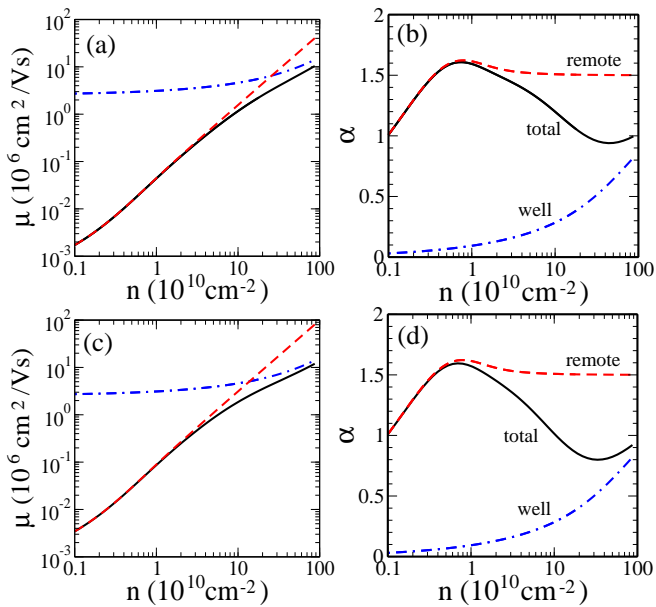}
\caption{(color online) 
(a) Hole mobility as a function of hole density ($n$) of
p-GaAs quantum well with a width $a=200$ \AA \; and the hole mass of
$m=0.4$. Here the background 
unintentional charged impurities with a density
$n_{i3D}=3 \times 10^{13} cm^{-3}$ are located inside the quantum well 
and 2D remote charged impurities with a density 
$n_i=8 \times 10^{11} cm^{-2}$ are located at $d= 800$ \AA \; from the
interface. In this figure
the black solid curve indicates the total mobility and blue dot-dashed
(red dashed) curve indicates the  
mobility limited by only background scattering (remote charged scattering). 
(b) The exponents ($\alpha$) for the corresponding mobilities in (a).
(c) Hole mobility with the same parameters of Fig.~\ref{fig:five}(a) except the
remote charged impurity density  
$n_i=4 \times 10^{11} cm^{-2}$.
(d) The exponents ($\alpha$) of corresponding mobilities of (c).
\label{fig:six}
}
\end{figure}

In Figs.~\ref{fig:five} and \ref{fig:six}, we show our 2D p-GaAs
results for hole-doped high-mobility GaAs quantum wells. Everything in
Figs.~\ref{fig:five} and \ref{fig:six} is identical except for using
two different hole effective mass values: $m_h = 0.3 m_0$
(Fig.~\ref{fig:five}) and $0.4 m_0$ (Fig.\ref{fig:six}) since the hole
effective mass in GaAs quantum wells is somewhat
uncertain\cite{eighteen}. The precise value of $m_h$ affects $q_{TF}
\propto m$, and thus determines the value of $q_s = q_{TF}/2k_F$,
leading to some difference between the results in Figs.~\ref{fig:five}
and \ref{fig:six}. For the holes, we show both the individual mobility
and exponent for each scattering process as well as the total mobility
and total exponent obtained by adding the two resistivities (or
equivalently, the two scattering rates arising from the two scattering
mechanisms). We deliberately refrain from showing the hole results
(Figs.~\ref{fig:five} and \ref{fig:six}) in the same format as the
electron results (Figs.~\ref{fig:one} and \ref{fig:two}) since they
would all look identical except for some changes in the numbers.


In Figs.~\ref{fig:five}/\ref{fig:six} (a) and (c) we show the
calculated 2D hole mobility $\mu(n)$ limited by 3D Coulomb scattering
($n_{i3D}$) and 2D remote Coulomb scattering ($n_i$) for
$a=200$ \AA \; with the only difference being $n_i = 8 (4)
\times 10^{11}$ cm$^{-2}$
respectively in Fig.~\ref{fig:five}/\ref{fig:six} (a) (c), showing
explicitly the quantitative importance of 2D remote Coulomb scattering
vis a vis 3D Coulomb scattering. The corresponding critical exponent
$\alpha(n)$, shown in Fig.~\ref{fig:five}/\ref{fig:six} (b) and (d)
respectively, is completely consistent with Table I with $\alpha$ for
remote scattering quickly reaching the asymptotic unscreened value of
3/2 as $2k_F d \gg 1$ and $\alpha$ for 3D Coulomb disorder increasing
slowly from the very strongly screened low density situation
($\alpha_{3D} \alt 0.1$ for $n < 2\times 10^{9}$ cm$^{-2}$) to $\alpha
\sim1$ for very high hole density ($n \sim  10^{12}$ cm$^{-2}$) where
screening weakens.

The total exponent in Figs.~\ref{fig:five} and \ref{fig:six} shows
complicated non-monotonicity as a function of carrier density since
the low-density (high-density) situation is more strongly affected by
remote 2D (background 3D) Coulomb scattering and the density dependent
crossover between the two scattering regimes is completely
nonuniversal depending precisely on the relative amounts of 2D and 3D
Coulomb disorder (i.e., on $n_{i3D}$, $n_i$, and $d$). The only
concrete statement one can make is that $\alpha(n)$ increases at low
carrier density generally to a value larger than unity whereas at
intermediate to high density it tends to stay below unity. Again, all
of these results are completely consistent with the asymptotic
exponents given in Table I (as long as various scattering mechanisms
with different exponents are combined together).

So far we have discussed our numerical results of
Figs.~\ref{fig:one}---\ref{fig:six} in terms of their consistency with
the theoretically analytically obtained critical exponents given in
Table I with the mobility exponent ($\mu \sim n^{\alpha(n)}$) $\alpha$
showing the expected behavior in the asymptotic density regimes of
$2k_F d \gg 1 \; (\ll 1)$ and $q_s \gg 1 \;  (\ll 1)$ as the case may
be. Remote 
scattering by 2D ionized dopants dominates transport at low density
($2k_F d < 1$) crossing over to background impurity scattering
dominated regime at higher density, leading to $\alpha > 1 \; (<1) $ at low
(high) density.

Now we discuss perhaps the most interesting aspect of our numerical
results in Figs.~\ref{fig:one}---\ref{fig:six}, which
appears to be at odds with our asymptotic theoretical analysis of
section III (and table I). This is the intriguing result that
$\alpha(n)$ due to remote 2D impurity scattering can actually exceed
the asymptotic unscreened 2D Coulomb scattering value of $\alpha =
3/2$. It is clear in Figs. \ref{fig:two}, \ref{fig:four},
\ref{fig:five}(b), \ref{fig:five}(d), \ref{fig:six}(b),
\ref{fig:six}(d) that there is a shallow maximum in $\alpha(n)$ at
some intermediate density where the numerically calculated mobility
exponent $\alpha >3/2 =1.5$ (and therefore the corresponding
conductivity exponent $\beta > 2.5$), which is remarkable since the
unscreened Coulomb scattering (applicable in the high-density regime
defined by $2k_F d \gg 1$ and/or $q_s \gg 1$) by remote impurities
produces  $\alpha =3/2$. This 
non-monotonic behavior of the exponent $\alpha(n)$ in the intermediate
density regime (neither high- nor low-density asymptotic regime
considered in Table I) with an exponent value larger than the
corresponding unscreened Coulomb exponent 3/2 is unexpected and highly
intriguing. We provide a theoretical explanation for this intriguing
nonmonotonic behavior of remote Coulomb scattering at intermediately
density with a mobility exponent exceeding the unscreened value of 1.5
in the next section.

\begin{figure}[t]
\includegraphics[width=.90\columnwidth]{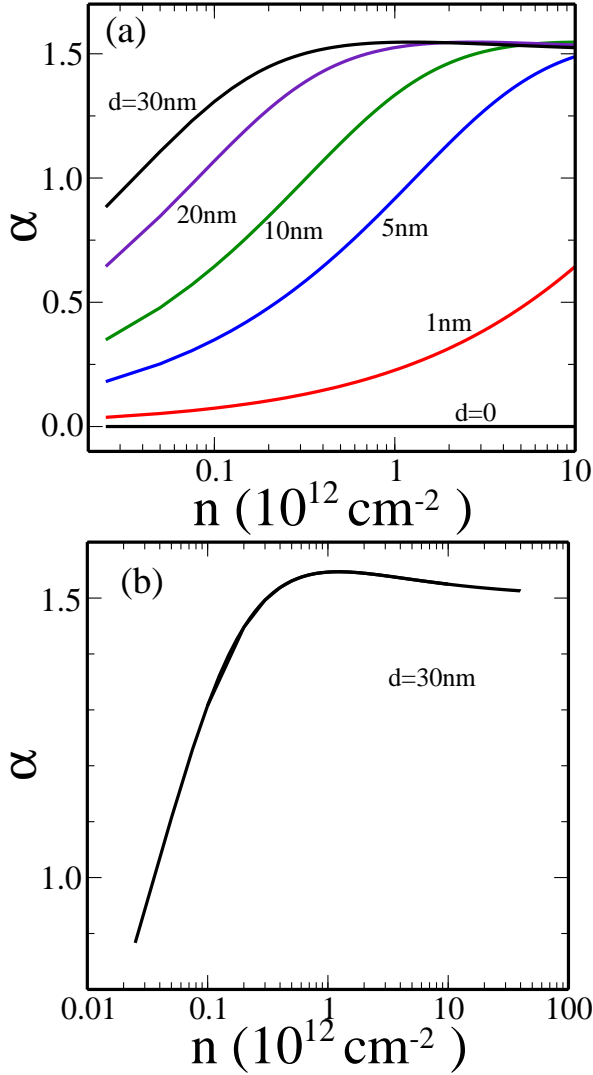}
\caption{(color online) 
(a) shows the calculated graphene
mobility exponent $\alpha(n)$ obtained from the numerically calculated
graphene mobility, $\alpha(n) =  d\ln \mu/d\ln n$, for various
location ($d$) of the 2D impurity layer in relation to the 2D
graphene layer. In (b) the exponent for $d=30$ nm shows a shallow
local maximum at $n \sim 10^{12}$ cm$^{-2}$.
\label{fig:seven}
}
\end{figure}

Before concluding this section on numerical results, we provide, for
the sake of completeness, our numerically calculated mobility exponent
for 2D graphene transport under the remote Coulomb scattering
situation. In Fig.~\ref{fig:seven} we show our calculated graphene
mobility exponent $\alpha(n)$ obtained from the numerically calculated
graphene mobility, $\alpha(n) =  d\ln \mu/d\ln n$, for various
locations ($d$) of the 2D impurity layer in relation to the 2D
graphene layer. We note that graphene is a strictly 2D system and
hence there is no quasi-2D form factor correction. It is clear that
$\alpha(n)$ goes asymptotically to the unscreened Coulomb value of
$3/2$ as $n$ (and therefore $k_F d$ increases) except for the trivial
$d=0$ case where $\alpha(n)=0$ (i.e., $\beta=1$) for all density as is
already well-known in the literature \cite{nine,sixteen} and is
well-verified experimentally \cite{nineteen}.

The calculated graphene mobility exponent $\alpha(n)$ shown in
Fig.~\ref{fig:seven} 
agrees completely with the analytical exponent values given in Table
I. (We mention again that in graphene the mobility exponent
$\alpha_{\mu}=\alpha$ and the relaxation rate exponent $\alpha_{\tau}$
differ with $\alpha=\alpha_{\mu} = \alpha_{\tau}-1/2$ and $\beta =
\alpha + 1 = \alpha_{\mu} + 1$ by definition whereas in 2D and 3D
parabolic system, where $\mu \propto \tau$, $\alpha_{\mu} =
\alpha_{\tau}= \alpha = \beta -1$.) We note that in graphene, for
impurities away from the 2D graphene plane (i.e., $d \neq 0$), the
asymptotic conductivity exponent $\beta$ for high carrier densities
($k_F d \gg 1$) is $3/2$ and thus $\sigma \propto n^{3/2}$ in graphene
layers dominated by far away Coulomb impurities.
An experimental verification of such a $\sigma \sim n^{3/2}$ behavior
in graphene due to Coulomb scattering by remote impurities will be a
direct verification of our theory.

\section{Nonmonotonicity of transport scaling}

We now theoretically discuss (and explain analytically) our surprising
numerical finding in section IV, not anticipated at all in the
asymptotic theory of section II or in any of the substantial earlier
literature on 2D transport, that the density scaling exponent $\alpha$
($\beta$) of 2D mobility (conductivity) has an intriguing
nonmonotonicity as a function of carrier density in the intermediate
density regime (in-between the asymptotic low and high density regimes
discussed in section III and tabulated in Table I).

We first note that the nonmonotonicity in $\alpha(n)$ arises from the
subtle fact that although the exponent $\alpha$ (or $\beta$) depends
only on one explicit external variable (namely, the carrier density
$n$), it depends theoretically on two independent dimensionless
variables $q_s = q_{TF}/2k_F$ and $d_0 = 2k_F d$ since in reality
there are two independent external variables in the problem: carrier
density ($n$) and the impurity location ($d$). The dependence on two
independent variables is the key feature allowing for the presence of
nonmonotonicity in $\alpha(n)$ as well as its maximum possible value
being larger than the unscreened exponent value $\alpha \rightarrow
3/2$. Indeed in the strict 2D limit with $d=0$ (see, e.g., the
graphene result in Fig.~\ref{fig:seven}), there is no maximum allowed
in $\alpha(n)$. This is true for both graphene and 2D parabolic system
-- in graphene, $\alpha(n)=0$ for all values of $n$ in the $d=0$ limit
whereas in strictly 2D parabolic system $\alpha(n)$ monotonically
increases from $\alpha=0$ in the low-density limit ($q_s \gg 1$) to
precisely $\alpha=1$ in the high density ($q_s \ll 1$) for $d=0$ as
one would expect theoretically (we have verified this strict 2D limit
with $d=0$ result explicitly numerically).

For $d\neq 0$ (i.e., $2k_F d \neq 0$), however, the behavior of
transport properties depends nontrivially on the variable $d_0 = 2k_F
d$, and there is no obvious theoretical reason why the low-density
($\alpha =0$ or 1 depending on strong or weak screening) and the
high-density ($\alpha = 3/2$ always) asymptotic limits must be the
lower and upper bounds on the exponent. In fact, as our numerics show,
and we establish theoretically below, $\alpha(n)$ does have a peak
(exceeding the unscreened $\alpha=3/2$ value) at an intermediate
density around $2k_F d \approx 1$.

As shown in our numerical results presented in the last section,
$\alpha(n)$ defined by $\mu \sim n^{\alpha(n)}$, can exceed the
unscreened exponent value $\alpha = 3/2$ reached in the asymptotic
high-density ($2k_F d \gg 1$) limit. To verify whether this finding is
a numerical artifact or real, it is sufficient to consider the
unscreened remote 2D Coulomb disorder in the strict 2D electron layer
limit where the free carriers and the charged impurities, both
confined in infinite zero-thickness 2D layers in the $x$-$y$ plane,
are separated by a distance $d$ in the $z$-direction. (This is the
model explicitly used in section III.) As $n \rightarrow 0$,
$\alpha(n) \rightarrow 1$ (unscreened $2k_F d \ll 1$ limit) whereas as
$n \rightarrow \infty$, $\alpha(n) \rightarrow 3/2$ (unscreened $2k_F
d \gg 1$ limit). The zero-density limit would be modified to $\alpha
=0$ if screening is included in the theory, but we are interested here
in the intermediate-density behavior, not the zero-density regime.

The scattering rate $\tau^{-1}$ in this case is given by [see
Eqs.~(\ref{eq:eq34}) and (\ref{eq:eq35})]:
\begin{equation}
\tau^{-1} = \frac{C}{n} \int_0^1 \frac{e^{-b x
    \sqrt{n}}}{\sqrt{1-x^2}} dx,
\label{eq:eq52}
\end{equation}
where we have shown the explicit density ($n$)-dependence everywhere
($C$ and $b$ are unimportant carrier density independent constants for our
purpose) and have used the fact that $k_F \sim \sqrt{n}$. We rewrite
Eq.~(\ref{eq:eq52}) as,
\begin{equation}
\tau^{-1} = \frac{C}{n}I(n),
\label{eq:eq53}
\end{equation}
where
\begin{equation}
I(n) = \int_0^1 \frac{e^{-b x \sqrt{n}}}{\sqrt{1-x^2}} dx.
\label{eq:eq54}
\end{equation}
It is obvious that $\tau^{-1}$ is a monotonic function of $n$
decreasing continuously with increasing density with no extrema
whatsoever since both $1/n$ and $\exp(-b x \sqrt{n})$ decrease with
increasing $n$ continuously.
This can be easily checked explicitly by showing that the equation
$d\tau^{-1}(n)/d n = 0$ has no solution. The monotonic decrease of
$\tau^{-1}(n)$ with $n$ simply implies that $\mu(n) \propto \tau(n)$
increases monotonically with increasing density as is obvious from our
numerical results in section IV: For Coulomb disorder, 2D mobility and
conductivity always increase with increasing density monotonically.

To figure out whether $\alpha(n) = d \ln \mu/ d \ln n$ has
non-monotonicity (or extrema) as a function of $n$, we must use
$\mu(n) \propto \tau(n)$, and write
\begin{equation}
\alpha = \frac{d \ln \mu}{d \ln n} = n \frac{d \ln \mu}{d n} = 1-
\frac{n}{I} \frac{d I}{dn},
\label{eq:eq55}
\end{equation}
where $I(n)$ is the integral defined by Eq.~(\ref{eq:eq54}). It is
straightforward, but messy, to show that the condition $d\mu/dn = 0$
with $\alpha(n)$ defined by Eq.~(\ref{eq:eq55}) has a solution at an
intermediate value of $n$, and thus $\alpha(n)$ has an extremum -- it
is still messier to show that $d^2 \alpha/dn^2 < 0$ at this extremum
so that $\alpha(n)$ has a maximum at an intermediate density as is
found numerically in section IV. For our purpose, however, it is much
easier to simply establish that the function $\alpha(n)$ defined by
Eq.~(\ref{eq:eq55}) approaches the high density $n \rightarrow \infty$
limit of $\alpha(n\rightarrow \infty) = 3/2$ from above, thus
definitively proving that the mobility exponent
$\alpha$ exceeds $3/2$ at some intermediate density (and thus must
have a maximum in the $0 < n < \infty$ or equivalently in the $ 1
\gg k_F d \gg 1$ regime).

As $n \rightarrow 0$, we have from Eq.~(\ref{eq:eq54}):
\begin{equation}
I(n) = \pi - b \sqrt{n},
\end{equation}
leading to 
\begin{equation}
\alpha(n\rightarrow 0) = 1 + b \sqrt{n}/\pi.
\end{equation}
As $n \rightarrow \infty$, we have:
\begin{equation}
I(n) = \frac{1}{b\sqrt{n}} \left [ 1 + \frac{1}{b^2 n} \right ],
\end{equation}
leading to 
\begin{equation}
\alpha(n\rightarrow \infty) = 3/2 + 1/2b^2 n.
\label{eq:eq59}
\end{equation} 
From Eqs.~(\ref{eq:eq34}), (\ref{eq:eq35}), (\ref{eq:eq52}), and
(\ref{eq:eq53}) we have: 
$ b = 2 \sqrt{2 \pi} d$, and thus $b \propto d$ is positive definite
(except for the $d=0$ explicitly left out here). We, therefore,
immediately conclude that $\alpha(n)$ approaches its asymptotic value
of $\alpha = 3/2$ for $n \rightarrow \infty$ from above with
$\alpha(n\rightarrow \infty) \approx \frac{3}{2} + \frac{1}{16\pi d^2}
\frac{1}{n}$, and the leading possible correction to the exponent in
the $n \rightarrow \infty$ limit is of $O(1/n)$ with a coefficient
$1/16 \pi d^2 \propto d^{-2}$. We also find that the correction to the
$n \rightarrow 0$ value of $\alpha(n=0) = 1$ is of $O(\sqrt{n})$ with
a positive coefficient of $2 \sqrt{2\pi} d/\pi \propto d$. Thus, we
now have a complete theoretical understanding of the intriguing (and
hitherto unexpected in the literature) numerical finding in Sec.~IV
that, although the mobility $\mu(n)$ itself is a monotonically
increasing function of increasing carrier density, its power law
exponent $\alpha(n)$ shows a maximum (around $2k_F d \sim 1$ in fact)
approaching the asymptotic high-density ($n \rightarrow \infty$;
$2k_Fd \gg 1$) value of $\alpha = 3/2$ from above, allowing
$\alpha(n)$ to be larger than 1.5 at some $d$-dependent value $\alpha
> 1.5$.

One feature of our theory presented in Eqs.~(\ref{eq:eq52}) --
(\ref{eq:eq59}) is 
worth mentioning and comparing with the numerical results of
Sec. IV. This is our finding in Eq.~(\ref{eq:eq59}) that the
asymptotic high-density ($n \rightarrow \infty$) exponent
$\alpha(n\rightarrow \infty) = 3/2$ is approached from above as
$\alpha(n \rightarrow \infty) = 3/2 + 1/(16 \pi d^2 n)$, implying that
the maximum possible values of $\alpha$, $\alpha_{\rm max}$, scales
approximately as $(d^2 n_{\rm max})^{-1} \propto (k_{Fm} d)^2$, where
$n_{\rm max}$ and $k_{Fm}$ are respectively the carrier density and
the corresponding Fermi wave vector at the maximum. This implies that
the maximum value $\alpha_{\rm max} \approx 1.7$ ($\approx 3/2 + 1/16
\pi$) is approximately independent of the value $d$ and of the carrier
effective mass with the value of the carrier density $n_{\rm max}$
(where the maximum occurs) scaling roughly as $n_{\rm max} \sim
d^{-2}$. This strong prediction is approximately consistent with our
numerical results -- in fact, $\alpha_{\rm max} \sim 1.7$ is clearly
independent of whether the system is a 2D electron or hole system and
of the precise value of $d$. In fact, $\alpha_{\rm max} \sim 1.7$
being
approximately independent of electrons/holes and the value of the
separation distance $d$ is a striking theoretical result which is
consistent with the full numerical results of Sec. IV.

\begin{figure}[t]
\includegraphics[width=.90\columnwidth]{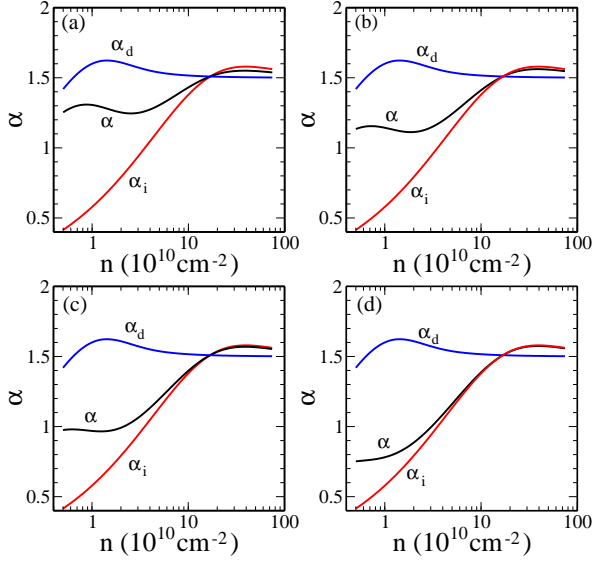}
\caption{(color online) 
The calculated mobility exponent as a function of density 
for 2D n-GaAs wells of thickness $a =
300$ \AA.  The mobility is calculated by incorporating both remote
impurity scattering with 2D 
impurity densities $n_{d}$ and separation $d$ and 
near impurity scattering with 2D impurity density $n_i$.
Here we use  $d=500$ \AA \; and   $n_i = 3 \times 10^8$ cm$^{-2}$ for
all figures, but different values of $n_d$, i.e.,  
(a) $n_d = 2\times 10^{10}$ cm$^{-2}$, (b) $n_d = 1\times 10^{10}$
cm$^{-2}$, (c) $n_d = 0.5\times 10^{10}$ cm$^{-2}$,
(a) $n_d = 0.22\times 10^{10}$ cm$^{-2}$.
The calculated exponents for the
individual scattering mechanisms $\alpha_d$ and $\alpha_i$ (for $n_d$
and $n_i$, respectively) and the total exponent
$\alpha$ are shown as a function of density. 
\label{fig:eight}
}
\end{figure}

Because of the striking nature of our finding that $\alpha_{\rm max}
\sim 1.7$ always (for remote impurity scattering) for 2D electron/hole
carrier systems, we carried out additional numerical calculations using
the realistic Boltzmann theory (including both quasi-2D finite
thickness and screening effects) for 2D n-GaAs wells of thickness $a =
300$ \AA \; (different from the case of $a=200$ \AA \; used in
Sec. IV) and incorporating both remote impurity scattering with 2D
impurity densities $n_{d}$ and separation $d$ and also (different
values of $n_d$ is used) with near impurity scattering with
2D impurity density $n_i$  with
$d=0$ (i.e. interface impurities). The calculated exponents for the
individual scattering mechanisms $\alpha_d$ and $\alpha_i$ (for $n_d$
and $n_i$, respectively) 
are shown in Fig.~\ref{fig:eight} where each panel corresponds to
different sets of values for $n_d$ and a fixed $n_i$. In each case, the
individual exponents $\alpha_d/\alpha_i$ as well as the total exponent
$\alpha$ are shown as a function of density. The exponent is extracted
from a full numerical evaluation of $\mu(n)$ and then using $\alpha(n)
= d\ln \mu(n)/d \ln n$, where the total exponent is extracted by
adding the two individual resistivities, i.e., $\mu^{-1} = \mu^{-1}_d +
\mu^{-1}_i$. The rather amazing fact to note in Fig.~\ref{fig:eight}
is that each individual exponent $\alpha_d$ and $\alpha_i$ has a
maximum value $\sim 1.7$, albeit the maximum for remote (near)
impurities occurring at low (high) carrier densities because the
effective $d$-values are much higher (lower) for remote (near)
impurities. (We note that $d=0$ for $n_i$ impurities still has an
effective $d$ value of roughly $a/2 \approx 150$ \AA \; whereas the
effective $d$ for the $n_d$ impurities is $d + a/2$ in each case.)
Figure \ref{fig:eight} is a striking direct numerical verification of
our theory.

Before discussing the same physics for graphene, which we do next, we
mention that including screening in the theory is straightforward (but
extremely messy). All we need to do is to modify Eq.~(\ref{eq:eq54})
to 
\begin{equation}
I(n) = \int_0^1 \frac{x^2 e^{-b x \sqrt{n}} dx}{(x+ c/\sqrt{n})^2
  \sqrt{1-x^2}},
\end{equation}
with $c=0$ giving the unscreened formula of
Eq.~(\ref{eq:eq52}). Inclusion of screening strongly affects the
low-density $n \rightarrow 0$ behavior, changing $\alpha(n \rightarrow
0)$ exponent to zero (for $c\neq 0$) from unity ($c=0$), but does not
affect the intermediate or high density behavior at all (as is obvious
from Table I where the $2k_F d \gg 1$ asymptotic results are
independent of the screening constant $q_s = q_{TF}/2k_F \propto
1/\sqrt{n}$). Since the extrema behavior of interest to us is not a
low-density phenomenon, our analysis based on Eq.~(\ref{eq:eq54}) is
appropriate (as is verified by its agreement with the full numerical
results).

Now, we consider the corresponding graphene case, which also has a
maximum 
in $\alpha(n)$ at some intermediate carrier density with $\alpha_{\rm
  max}$ ($>1.5$) being larger than the corresponding infinite density
unscreened exponent value of 3/2 (see Fig.~\ref{fig:seven}). For
graphene, with the charged impurities located in a 2D layer a distance
$d$ from the graphene layer, the scattering rate $\tau^{-1}$ is given
by [see Eq.(\ref{eq:eq41})]:
\begin{equation}
\tau^{-1} = A \sqrt{n} \int_0^1 dx \frac{x^2 \sqrt{1-x^2}}{(x+q_s)^2}
  e^{- \tilde{b} x \sqrt{n}},
\label{eq:eq62}
\end{equation}
where $A$, $\tilde{b} = 2 \sqrt{\pi} d$ are density-independent
constants and $q_s = 4 e^2/\kappa \hbar v_0$ is also density
independent. Direct expansion for Eq.~(\ref{eq:eq62}) in the low ($n
\rightarrow 0$) and high ($n \rightarrow \infty$) limits give:
\begin{equation}
\tau^{-1}(n \rightarrow 0) = \frac{A_0}{\sqrt{n}} \left ( 1- \frac{16
    d}{3 \pi} \sqrt{\pi n} \right ),
\end{equation}
and
\begin{equation}
\tau^{-1}(n \rightarrow \infty) =
\frac{A_{\infty}d}{(k_Fd)^2(q_{TF}d)^2} \left ( 1- \frac{6}{q_{TF} d}
\right ),
\end{equation}
where $A_0$ and $A_{\infty}$ are constants independent of $n$ and
$d$. 
For graphene $\mu(n) \sim \tau/\sqrt{n}$, and therefore, we get for $n
\rightarrow 0$
\begin{equation}
\mu(n) \sim \left ( 1 + \frac{16d}{3\pi} \sqrt{n} \right ),
\end{equation}
and for $n \rightarrow \infty$
\begin{equation}
\mu(n) \sim n^{3/2} d^3 \left ( 1 + \frac{3}{2 r_s d \sqrt{\pi n}}
  \right ),
\label{eq:eq66}
\end{equation}
where $r_s = e^2/(\kappa \hbar v_0)$ is the so-called graphene fine
structure constant.

Eqs.~(\ref{eq:eq62}) -- (\ref{eq:eq66}) imply that the mobility
$\mu(n)$ starts at low density ($n \rightarrow 0$) with $\alpha = 0$,
but with a leading order correction going as $O(\sqrt{n})$ with a
positive sign. For large $n$ (i.e., $k_F d \gg 1$), $\alpha(n
\rightarrow \infty)$ becomes 3/2, but has a positive leading order
correction of $O(1/\sqrt{n})$. This immediately implies that
$\alpha(n)$ must have a local maximum at some intermediate density
with $\alpha$ being only logarithmically larger than the asymptotic
($n \rightarrow \infty$) value of 3/2. This conclusion is consistent
with numerical results presented in Fig.~\ref{fig:seven}.
We note that the maximum in $\alpha(n)$ for graphene is much shallower
and weaker than in the 2D parabolic system.

\section{Experimental Implications}

We now discuss the possible experimental relevance of our
theoretical findings in this section. The fact that the mobility
$\mu(n)$ or the conductivity $\sigma(n) = n e \mu$ of 2D carrier
systems shows a density scaling behavior with $\mu \sim n^{\alpha(n)}$
and $\sigma \sim n^{\beta(n)}$ with $\beta = \alpha + 1$ has been
known for a long time in the experimental 2D transport literature
\cite{ten}. Although our current work is purely theoretical, focusing entirely
on the fundamental questions of principle involving the detailed
behavior of the exponent $\alpha(n)$ and $\beta = \alpha + 1$ for
various types of disorder affecting transport properties, we believe
that it is appropriate for us to comment on experiments, making
connection with the existing data in the literature as well as making some
concrete predictions for future experiments (particularly in the
context of our unexpected finding of a maximum in $\alpha(n)$ with an
almost universal value of 1.7 for 2D semiconductor quantum wells).

We first summarize the serious difficulties in making direct
quantitative connection or comparison between experiment and theory in
2D semiconductor systems with respect to low-temperature
disorder-limited transport properties. In fact, these caveats apply to
all disorder-limited transport properties in all systems, not just to
2D transport in semiconductor quantum wells. The key problem is that
the detailed nature of disorder (either qualitative or quantitative)
in a sample is never precisely known from independent measurements ---
in the context of our transport theory in this paper, relative amounts
of 2D near and remote Coulomb disorder, 3D Coulomb disorder, and
short-range disorder are simply not known. Thus, a quantitative or
even a qualitative theory for calculating the mobility or the
conductivity of a given sample exists only as a matter of principle,
but not in practice since the details of the underlying disorder
contributing to resistive scattering are apriorie unknown and often are
figured out indirectly based on quantitative comparisons between
transport experiment and theory. The situation is worsened by the fact
that the relative magnitudes of various independent scattering
mechanisms vary strongly with carrier density -- for example, Coulomb
disorder weakens with increasing carrier density making short-range
scattering relatively stronger at high carrier density. Thus, all 2D
semiconductor transport would eventually be dominated by short-range
scattering (e.g., interface roughness, alloy disorder) at very high
carrier density (where $\alpha \approx 0$ in 2D systems), and the only
question is how high in density one must go to reach this asymptotic
zero-range disorder limited regime where Coulomb disorder has
virtually been screened out. This, of course, is completely
nonuniversal and depends entirely on the relative amount of Coulomb
impurities and short-range disorder in particular samples! This
discussion shows that in 2D semiconductor we have $\alpha(n
\rightarrow 0) =0$ and $\alpha (n \rightarrow \infty) = 0$ purely
theoretically with the zero-density limit and the infinite-density
limit being dominated by completely screened Coulomb disorder ($k_F d
\ll 1$, $q_s \gg 1$) and zero-range disorder, respectively, although
these strict theoretical limits are unlikely to apply to real samples
at any finite carrier density.

The difficulty of applying the pristine theory to specific
experimental situations 
is obvious from our numerical results presented in Figs.~\ref{fig:five}
and \ref{fig:six} (for 2D holes) and Fig.~\ref{fig:eight} (for 2D
electrons). In each case, the mobility exponent $\alpha$ for
individual near and far Coulomb impurities follows our theoretical
prescription perfectly with the expected low- and high-density
exponents agreeing with the results of Table I, but the total exponent
$\alpha$ (which is the only one relevant for the experimental data)
may not follow any well-defined pattern and could vary strongly
depending on the relative strengths of near and far Coulomb disorder,
showing strong nonmonotonicity (Figs.~\ref{fig:five} and
~\ref{fig:six}) or weak/no nonmonotonicity (Fig.~\ref{fig:eight}).
This reinforces the point made earlier by us that the universal
density scaling behavior in transport applies only to individual scattering
mechanism with the overall transport being dominated by crossover
behavior is generically nonuniversal due to the existence of several
different operational scattering processes.

In spite of the above serious caveats arising from our ignorance about
the underlying disorder contributing to resistive scattering
mechanisms, some general statements can be made about the implications
of our theory to experimental data. We discuss this below.

(i) For very dirty (and low-mobility) samples, the background Coulomb
disorder arising from the unintentional charged impurities in the
quantum well should dominate transport properties, leading to
$\alpha_e \sim \alpha_h \sim 0.5$ for a wide range of intermediate
densities.
(ii) When transport is limited by remote dopants, which would always
be true in modulation-doped samples for $k_F d < 1$, $\alpha_e \sim
1.5$ and $\alpha_h \sim 1-1.5$ depending on the hole effective mass,
but $\alpha_e/\alpha_h$ will decrease with increasing density as $k_F d
> 1$ regime is reached.
(iii) For modulation doped structures with $k_F d \gg 1$, background
disorder again dominates at intermediate density giving $\alpha_e \sim
\alpha_h \sim 0.5 - 1$.

The above situation seems to describe the existing experimental
situation for 2D quantum well transport reasonably well as discussed
below.
Focusing on specific experimental results in the literature in the
context of our transport theory, we make the following remarks
discussing some specific experimental publications in 2D GaAs based
electron and hole systems.

(1) In Ref.\onlinecite{twenty}, the measured $\alpha_e \approx 0.6-0.7$
in n-GaAs 2D system with no intentional remote dopants (the sample is
a gated undoped sample) in the density range $\sim 10^{10} - 10^{11}$
cm$^{-2}$ (i.e., $q_s \agt 1$; $2k_F d < 1$) agrees quantitatively
with our theoretical results given in Figs.~\ref{fig:three},
\ref{fig:four}, and \ref{fig:eight} with background 2D and 3D
unintentional charged impurities being the main disorder mechanism as
expected for an undoped 2D system.

(2) In a similar gated undoped 2D p-GaAs sample Manfra {\it et al.}
\cite{twentyone} find $\alpha_h \sim 0.7$ for density $\agt 10^{10}$
cm$^{-2}$, again agreeing with our results given in
Figs.~\ref{fig:three} and \ref{fig:four} for background scattering.

(3) In Harrell {\it et al.} \cite{twentytwo} gated undoped n-GaAs 2D
samples, $\alpha \approx 0.6$ was found for $n \agt 10^{11}$ cm$^{-2}$
and $\alpha \approx 0.33$ was found for $n < 5 \times 10^{10}$
cm$^{-2}$. This is both quantitatively and qualitatively consistent
with our numerical findings in Figs.~\ref{fig:three}, \ref{fig:four},
and \ref{fig:eight}, where scattering by background charged impurities
in the layer leads to $\alpha \approx 0.3-0.7$ in the $n=10^{10} -
10^{11}$ cm$^{-2}$ density range with $\alpha(n)$ decreasing with
decreasing carrier density.

(4) Melloch {\it et al}. \cite{twentythree} found $\alpha \approx 0.6
- 0.7$ for $n >10^{11}$ cm$^{-2}$ which is consistent with our
background impurity scattering results.

(5) Pfeiffer {\it et al.} studied \cite{twentyfour} modulation-doped
high-mobility 2D GaAs electron systems \cite{twentyfour} obtaining
$\alpha \sim 0.7$ around $n\sim 3 \times 10^{11}$ cm$^{-2}$ for
modulation-doped structures ($d=1000-2000$\AA) with $\mu \agt 10^7$
cm$^2$/Vs. Again, remote impurity scattering is completely ineffective
here because $k_F d \gg 1$ rendering mobility limited only by remote
impurity scattering to be around $10^8$ cm$^2$/Vs according to our
numerical calculations. The dominant scattering mechanism in this
sample is by background unintentional charged impurities, leading to
$\alpha \approx 0.7$ around $n \sim 3 \times 10^{11}$ cm$^{-2}$
according to our Fig.~\ref{fig:four}, which is in precise agreement
with the date of Pfeiffer {\it et al.} \cite{twentyfour}.

(6) In a similar high-mobility modulation-doped 2D n-GaAs sample,
Shayegan {\it et al}. \cite{twentyfive} found $\alpha \approx 0.6$ in
samples with $\mu \approx 10^6$ cm$^2$/Vs for $n \alt 10^{11}$
cm$^{-2}$ with the spacer thickness $d = 1000-2000$ \AA. Again, remote
scattering by the intentional dopants is ineffective as a resistive
scattering mechanism here with the dominant scattering 
being by unintentional background
impurities in the GaAs quantum well.From Fig.~\ref{fig:four} of our
presented results, we find $\alpha \approx 0.6$ for $n \alt 10^{11}$
cm$^{-2}$ in agreement with the experimental finding of Shayegan {\it
  et al.} \cite{twentyfive}.

(7) Most of the high-mobility experimental samples discussed above are
dominated by the background unintentional charged impurities in the 2D layer
itself leading to $\alpha < 1$ by virtue of the fact that the
intentional dopants introduced for modulation doping are rather far
away in these high quality samples (this is a generic feature of all
high mobility 2D samples with $\mu > 10^6$ cm$^2$/Vs where $\alpha <1$
prevails by virtue of the background disorder being dominant). By
contrast, early work on modulation-doped 2D samples invariably had
lower values of $d$ and achieved much lower mobility $\mu < 10^6$
cm$^2$/Vs. Such samples are almost always dominated by remote
scattering by the intentionally introduced dopants, leading to
$\alpha$ values typically exceeding unity as our theory predicts. As a
typical example, we consider the work of Hirakawa and Sakaki
\cite{twentysix} on modulation-doped 2D n-GaAs samples with $\mu \sim
10^4 - 5 \times 10^5$ cm$^2$/Vs for $d\approx 0-180$ \AA \; in the
$n\approx 10^{11} - 5 \times 10^{11}$ cm$^{-2}$ density
range. Assuming transport to be limited entirely by the intentional
ionized dopants in the modulation layer (i.e., no background disorder
scattering), our results of Fig.~\ref{fig:one} predict $\alpha \approx
1$ for $d=0$ and $\alpha \approx 1.1 - 1.3$ for $d=100$ \AA. Hirakawa
and Sakaki reported \cite{twentysix} $\alpha \approx 1.1-1.3$ for
$d\approx 0-100$ \AA \; in essential agreement with our theory. In
addition, these authors reported $\alpha \approx 1.7$ for $d\approx
200$ \AA, an anomalous mobility exponent (i.e. $d > 3/2$) which has
remained unexpected in the literature for more than 25 years. Our
current work provide a definitive explanation for $\alpha \approx 1.7$
as arising from the maximum in $\alpha(n)$ for remote scattering as is
apparent in Fig.~\ref{fig:one}. We note that $k_F d \sim 1-2$ for $n
\sim 10^{11}$ cm$^{-2}$ and $d \sim 100 - 200$ \AA, and thus our
theory predicts $\alpha \sim 1.7$ for $d \approx 200$ \AA \; in the
Hirakawa-Sakaki experiment \cite{twentysix}. We believe that the
experimental finding of $\alpha \sim 1.7$ by Hirakawa and Sakaki is a
direct verification of our intriguing prediction of $\alpha > 3/2$ for
$k_F d \agt 1$ in transport dominated by remote scattering.

(8) Finally, we discuss some recent unpublished experimental work by
Pfeiffer and West \cite{twentyseven} who, motivated by our theoretical
work, carried out low-temperature transport measurements in a series
of high-quality (i.e. ultrapure GaAs with very little background
disorder due to unintentional impurities) MBE-grown 2D n-GaAs samples
with variable values of $d$. Since these experiments were performed
with the specific goal of checking 
our low-temperature 2D transport theory predictions, Pfeiffer and West
made undoped gated samples of highest quality with little background
disorder and a nominal low-temperature mobility of $\mu > 10^7$
cm$^2$/Vs. Then, they systematically introduced charged impurities at
specific separation ($d$) from the 2D layer by inserting carbon atoms
in the GaAs layer ($d=0$) or in AlGaAs barrier layer ($d\neq
0$). First they explicitly verified that the introduction of different
amounts of impurity centers without changing $d$ only affects the
2D mobility through the expected $n_i \mu$ scaling behavior (i.e. $\mu
\propto n_i^{-1}$) without changing $\alpha(n)$ by changing the
inserted carbon atoms by a factor 5 keeping $d$ fixed (which
changed the 2D mobility by a factor of 5 without changing the exponent
$\alpha(n)$ in the same carrier density range). Thus, they measured
$\alpha(n)$ for $d=0$ and $d=150$ \AA \; finding $\alpha(n) \approx
0.8$ and 1.8 respectively for $n \sim 10^{11}$ cm$^{-2}$. Our
calculated $\alpha$($n \approx 10^{11}$ cm$^{-2}$)$\approx 0.8$ for
$d=0$ in Fig.~\ref{fig:two} in perfect agreement with the experimental
data. For $d=150$ \AA, $k_Fd \approx 1$ for $n \approx 10^{11}$
cm$^{-2}$, and we predict $\alpha(n) \approx \alpha_{\rm max} = 1.7$
in this situation which compares well with the experimental exponent
of 1.8. Thus, this experimental investigation of our theoretical
predictions appears to have strikingly verified our finding that
$\alpha > 3/2$ in the $k_F d \sim 1$ intermediate density regime where
the shallow maximum occurs in $\alpha(n)$ in our theory.

(9) Before concluding our discussion of experimental implications of
our theory we describe the very recent 2D hole transport data in
high-mobility p-GaAs systems by the Manfra group
\cite{twentyeight,twentynine}. These 2D p-GaAs samples with hole
mobility $>2\times 10^6$ cm$^2$/Vs are the world's highest mobility
hole samples ever, and taking into account the effective mass
difference ($\sim$ a factor of 5) between GaAs electrons and holes
compare favorably with the best ($\sim 15\times 10^6$ cm$^2$/Vs)
available 2D electron mobilities. The main finding of the work
\cite{twentyeight} is that the mobility exponent $\alpha(n)$ increases
from 0.7 at high hole density ($\sim 10^{11}$ cm$^-2$) to 1.7 at low
density ($\agt 10^{10}$ cm$^{-2}$) for modulation doped sample with
$d=800$ \AA. Since the mobility remains high throughout ($\gg 10^5 $
cm$^2$/Vs), localization effects should not be playing a role. We
therefore believe that the experimental finding of Watson {\it et al.}
\cite{twentyeight} is a direct confirmation of the $\alpha(n)$
behavior for 2D holes presented in our Figs.~\ref{fig:five} and \ref{fig:six},
where the total calculated $\alpha(n)$ increases monotonically as the
2D hole density decreases from $n\sim 10^{11}$ cm$^{-2}$  to $n \sim
10^{10}$ cm$^{-2}$. In fact, even the explicit $\alpha$-values
measured by Watson {\it et al.} agree well with our 2D hole
theoretical results presented in Figs.~\ref{fig:five} and
\ref{fig:six} with $\alpha$ ($n \sim 10^{10}$ cm$^{-2}$) increasing
above the unscreened $\alpha=3/2$ value reaching essentially the
measured value of $\alpha \sim 1.7$ around $n \agt 10^{10}$
cm$^{-2}$. Thus, our theory provides a qualitative explanation of the
Watson {\it et al}. experimental results including the surprising
finding of the intermediate-density $\alpha$ being around 1.7 ($>
1.5$). 

(10) We conclude this section on the experimental relevance of our
theory by discussing graphene briefly. There has been substantial
research activity on studying the density dependent graphene
conductivity \cite{nine,thirty} which is well beyond the scope of our
current work and has already been covered elsewhere. It is known that
scattering by near random charged impurities located on the surface of
the graphene layer or at the graphene-substrate interface 
leads to a $\sigma(n) \propto n$ (i.e., $\mu(n) \sim$ constant) in the
intermediate density ($k_F d < 1$) region, and in the high-density
regime $\sigma(n)$ becomes sublinear most likely because of
short-range defect scattering which gives $\sigma(n)\sim$ constant
(i.e., $\mu \sim 1/n$) -- see Table I for details. Theory predicts
(Table I) that for $k_F d \gg 1$, i.e., remote Coulomb disorder,
$\mu(n)$ and $\sigma(n)$ should cross-over to $\mu(n) \sim \sqrt{n}$
and $\sigma(n) \sim n^{3/2}$ in graphene. This clear prediction could
be verified by putting an impurity layer (e.g. a SiO$_2$ film) at
various values of $d$ from the graphene layer and measuring
$\sigma(n)$ to check if the low-density ($2k_F d \ll 1$) $\sigma(n)
\sim n$ behavior indeed crosses over to the high density $\sigma(n)
\sim n^{3/2}$ behavior as we predict theoretically. In graphene, with
a valley degeneracy of 2, $2k_F d \approx 2.5 \tilde{d}
\sqrt{\tilde{n}}$, where $\tilde{d} = d/1000$\AA, and $\tilde{n} =
n/10^{10}$cm$^{-2}$. Hence, for $n=10^{12}$ cm$^{-2}$ and $d=100$ \AA
\; $2k_F d \approx 2.5$. Thus, the condition $2k_F d \gg 1$ would
require an impurity layer at $d \approx 1000$ \AA \; (with consequent
very weak Coulomb disorder scattering) which may lead to the
complication that the subsequent graphene resistivity will be entirely
dominated by any underlying short-range disorder (with $\sigma \sim
n^0$), masking any $\sigma \sim n^{3/2}$ behavior arising from charged
impurity scattering. One possibility would be to put suspended
graphene near a thick layer (of thickness $L$) of disordered substrate
with a 3D charged impurity distribution, which would lead to [see
  Eqs.~(\ref{eq:eq41})--(\ref{eq:eq43}):
\begin{equation}
\tau^{-1} = \frac{N_i}{\pi \hbar^2 v_0 k_F^2} \left ( \frac{2\pi
  e^2}{\kappa} \right )^2 \int_0^1 dx \frac{x\sqrt{1-x^2}}{(x+q_s)^2}
  \left [ 1-e^{-2L_0 x} \right ]
\end{equation}
with $N_i$ being the 3D impurities density in the impurity layer and
$L_0 = 2k_F L$, giving $\sigma(n) \sim n^{3/2}$ for $k_F L \gg 1$ and
$\sigma(n) \sim n$ for $k_F L \ll 1$. Such a 3D impurity layer
underneath graphene may manifest the superlinear $\sigma(n) \sim
n^{3/2}$ conductivity behavior predicted by the theory, but observing
the shallow maximum in the exponent $\beta$/$\alpha$ for $k_F d \sim
1$ may still be difficult.

\section{other effects}

In this section, just before our conclusion in the next section, we
discuss ``other effects'' completely left out of our theoretical
considerations which may compromise and complicate direct quantitative
comparisons between our theory and experiment although we believe that
our theoretical conclusions should apply generically to transport in
high-mobility 2D semiconductor systems at low enough temperatures.

First, phonon effects are neglected in the theory since we explicitly
consider the $T=0$ situation (in practice, $T=50-300$ mK is the
typical low-temperature experimental situation mimicking the $T=0$
theoretical situation). For consistency between theory and experiment,
the transport data must therefore be taken at a fixed temperature
$T<T_{BG}$, where $T_{BG}$ is the so-called Bloch-Gr\"{u}neisen
temperature, so that acoustic phonon scattering contribution to the
resistivity is negligible compared with disorder scattering even in
the high-mobility 2D semiconductor structures under consideration in
this work. (Optical phonon transport is of no relevance for
low-temperature transport since $k_BT \ll \hbar \omega_{LO}$ is
explicitly satisfied as $\hbar \omega_{LO} > 100$ K typically.)
$T_{BG}$ is given either by the Debye temperature (for 3D metals) or
by the energy of the acoustic phonons with $2k_F$-wave vector (for 2D
semiconductors), whichever is lower. We therefore have $k_B T_{BG} = 2
\hbar k_F v_{ph}$ where $v_{ph}$ is the relevant acoustic phonon
velocity in the material. Putting in the appropriate sound velocity
($v_{ph}$), we get $T_{BG} \approx 2 \sqrt{\tilde{n}}$ K; $10
\sqrt{\tilde{n}}$ K in 2D GaAs and graphene, respectively, where
$\tilde{n}$ is measured in units of $10^{10}$ cm$^{-2}$. Thus, down to
carrier density $n \sim 10^9$ cm$^{-2}$, it is reasonable to ignore
phonon effects in transport at $T \approx 100$ mK. Acoustic phonon
scattering has been  considered elsewhere in the literature \cite{eight}.

Second, we have used the Born approximation in calculating the
scattering time and the RPA in calculating the screened Coulomb
disorder throughout. Both of these approximations surely
become increasingly quantitatively inaccurate at lower carrier density
although they should remain qualitatively valid unless there is a
metal-insulator transition (obviously, our Drude-Boltzmann transport
theory would not apply at or below any metal-insulator transition
density). RPA screening theory becomes increasingly quantitatively
inaccurate as carrier density (or the corresponding $r_s \sim
n^{-1/2}$ parameter) decreases (increases), but there is no
well-accepted systematic method for incorporating low-density
electronic correlation effects going beyond RPA screening. Including
low-density correlation effects in Hubbard-type local field theories
do not change any of our qualitative conclusion. As for multiple
scattering effects \cite{gold} beyond Born approximation, they
typically lead to higher-order corrections to the resistivity so that
$\rho \propto n_i$, where $n_i$ is the impurity density, is no longer
valid and one must incorporate high-order nonlinear corrections to the
resistivity going as $O(n_i^2)$ and higher. These nonlinear
multiscattering corrections become quantitatively important for $n
\alt n_i$, and can be neglected for $n > n_i$ regime, which is of our
main interest in this paper. Our neglect of multiscattering
corrections beyond Born approximation is consistent with our neglect
of strong localization effect, both of which will become important in
the very low carrier density regime ($n<n_i$) where the
Drude-Boltzmann theory becomes manifestly inapplicable.

Third, we ignore all nonlinear screening effects, which have been much
discussed in the recent graphene literature \cite{adam} where charged
impurity induced inhomogeneous electron-hole puddle formation plays an
important role at low carrier density, since they are important only
at very low carrier density ($n<n_i$) where our whole Boltzmann
theoretical approach becomes suspect any way. For the same reason we
also do not take into account any scattering-screening
self-consistency effect \cite{dassarma} which may also become
important at very low carrier density (again, for $n<n_i$).

Fourth, we ignore any possible spatial correlation effects among
impurity locations, assuming the disorder to arise from completely
uncorrelated random impurity configurations. If the impurities are
spatially correlated, it is straightforward to include the correlation
effect in the Boltzmann transport calculation by simply multiplying
the disorder potential term in Eq.~(\ref{eq:eq7}) by the corresponding
structure factor $s(q)$ for the impurity distribution, i.e., by
writing $|V_q|^2 \rightarrow |V_q|^2s(q)$ in Eq.~(\ref{eq:eq7}) where
\begin{equation}
s(q) = \frac{1}{n_i} \left | \sum_{i=1}^{n_i} e^{-i {\bf q \cdot
    r_{i}}}
  \right |^2 - n_i \delta_{q0},
\end{equation}
where ${\bf r_i}$ denotes the position of each impurity and the sum
going over all the impurities. This, if experimental information about
impurity correlations exists, it is then straightforward to include
such spatial correlation effects in the Boltzmann transport
calculations, as has indeed been done in both 2D GaAs\cite{cor_gaas} and
graphene \cite{correlation}. Since intrinsic impurity correlation information is
typically unavailable, our model of uncorrelated random disorder seems
to be the obvious choice from a theoretical perspective since it
involves only one (the impurity density $n_i$) or two ($n_i$ and the
impurity location $d$) unknown parameters (and most often $d$ is known
from the modulation doping setback distance and is thus not a free
parameter) whereas including impurity spatial correlations would
invariably involve the introduction of more unknown free parameters
making the theory of dubious theoretical relevance. Since our interest
in this paper is not an absolute calculation of the conductivity or
mobility (and hence $n_i$ typically drops out of our theory if only
one dominant scattering mechanism is operational), but obtaining the
universal density dependence of conductivity, it is important to
mention that the main effect of impurity correlation is to suppress
the effect of $n_i$ on the resistivity without much affecting the
carrier density dependence particularly for $n>n_i$ regime of our
interest. We mention that any intrinsic impurity correlations actually
increase the value of mobility from our uncorrelated random disorder
theory and thus compensate to some extent the suppression of mobility
arising from some of the effects discussed above.

Fifth, for our $T=0$ theory to be strictly applicable to the
experimental data, the experimental temperature $T$ should satisfy the
condition $T \ll T_F$, where $T_F \equiv E_F/k_B$ is the corresponding
Fermi temperature of the system. Using the known dependence of $E_F$
on the 2D carrier density $n$ we see that this implies $T (K) \ll 4.2
\tilde{n}$ for 2D n-GaAs, $T(K) \ll 1 \tilde{n}$ for 2D p-GaAs, and
$T(K) \ll 150 \sqrt{\tilde{n}}$ for graphene where $\tilde{n} =
n/10^{10}$cm$^{-2}$. Thus, low-temperature experiments carried out at
$T=100$ mK satisfy our $T=0$ theoretical constraint very well down to
$10^9$ cm$^{-2}$ carrier density (except for 2D p-GaAs hole system
where the density cut off is perhaps $5\times 10^9$ cm$^{-2}$). Our
direct numerical calculations at finite temperatures (not shown in
this paper) show that our calculated exponents $\alpha(n)$ and
$\beta(n) = \alpha +1$ at $T=0$ continue to be quantitatively accurate
upto $T \agt T_F$ as long as the experimental data for density
dependence are taken at a fixed temperature. Thus our theory and
numerics for $\alpha(n)$ and $\beta(n)$ are quite robust against
thermal effects as long as phonons are unimportant (i.e., $T < T_{BG}$
is satisfied).

We conclude this section of ``other effects'' by noting that the most
important drawback of our RPA-Drude-Boltzmann theory is that it may
fail systematically at low carrier density ($n \alt n_i$)
where important physical effects (which are difficult to treat
theoretically) such as strong localization, metal-insulator
transition, nonlinear screening, inhomogeneous puddle formation,
multiscattering corrections, screening-scattering self-consistent,
etc. may all come into play making both our theory inapplicable to the
experimental situation. We do, however, anticipate our theory to be
applicable to very low carrier densities ($n \agt 10^9$ cm$^{-2}$) in
ultra-high mobility 2D GaAs and graphene systems where $n_i \alt 10^8$
cm$^{-2}$ typically. A convenient experimental measure of the
applicability of our theory is looking at the dimensionless quantity
'$k_F l$' where $l$ is the elastic mean free path defined by $l \equiv
v_F \tau$ where $\tau$ and $v_F$ are respectively the transport
relaxation time and Fermi velocity. We find that $k_F l = 4.14
\tilde{n} \tilde{\mu}$ for 2D GaAs systems, where $\tilde{n} =
n/10^{10}$cm$^{-2}$ and $\tilde{\mu} = \mu/(10^6$ cm$^2$/Vs) whereas
$k_Fl=0.2 \tilde{n} \tilde{\mu}$ for graphene. As long as $k_F l > 1$,
our Drude-Boltzmann theory should be valid qualitatively and we
therefore conclude that the theory remains quantitatively accurate for
$n \agt 10^{10}$ cm$^{-2}$ ($10^{11}$ cm$^{-2}$) for 2D GaAs
(graphene) systems in high-quality/high-mobility samples. Thus, there
is a large range of carrier density ($10^{10}-10^{12}$ for 2D GaAs and
$10^{11} - 10^{13}$ for graphene) where our predictions for universal
density dependence can be experimentally tested through
low-temperature transport measurements.

The fact that all the ``other effects'' left out of our theory affects
transport at low carrier densities indicates that our predicted
density scaling of conductivity will systematically disagree with the
experimental data at lower carrier densities. Of course, ``lower''
density is a relative term, and the dimensionless quantities such as
$n/n_i$ and $k_F l$ are the appropriate quantities to define the
regime of validity of our theory. As $n/n_i$ and/or $k_F l$ (or even
$T_F/T$ or $T_{BG}/T$) become smaller, the Drude-Boltzmann theoretic
predictions become increasingly unreliable. Nevertheless, the theory
remains predictive down to $10^{10}$ cm$^{-2}$ carrier density (or
lower) in high quality (i.e. low values of $n_i$) GaAs samples at low
temperatures ($T \approx 100$ mK).

\section{discussion and conclusion}

We have developed a detailed quantitative theory for the
density-dependence of the zero-temperature conductivity (or
equivalently mobility) of (mainly) 2D and 3D electron and hole
metallic systems assuming transport to be limited by (mainly Coulomb)
disorder scattering within the semiclassical Drude-Boltzmann transport
theory. We neglect all quantum interference (hence localization)
effects as well as interaction effects (except for the carrier
screening of the bare impurity Coulomb disorder, which is an essential
qualitative and quantitative ingredient of our theory) assuming them
to be small since our interest is the density dependence (rather than
the temperature dependence) of transport properties at low fixed
temperatures in 
high-mobility ($k_F l \gg 1$ where $l$ is the elastic mean free path)
samples.

We have systematically considered 3D and 2D doped (n- and p-)
semiconductor systems as well as 2D graphene but the primary focus has
been on n-GaAs and p-GaAs quantum well based 2D electron or hole
systems, mainly because these systems continue to be of great interest
in physics and because the carrier density can easily be tuned in such
high-mobility 2D semiconductor systems, and the mobility is dominated
by Coulomb disorder at low temperatures. We have taken into account
both long-range Coulomb disorder from charged impurities and
zero-range disorder arising from possible non-Coulombic short range
scatterers. The primary focus has been the Coulombic disorder since
this is the main low-temperature resistive scattering
mechanism in semiconductors. Instead of discussing the 
nonuniversal values of $\mu$ (and $\sigma$), which depend
\cite{seventeen} on the actual impurity content, we focus on the
universal power-law density scaling of transport properties:
$\sigma(n) \sim n^{\beta(n)}$ and $\mu(n) \sim n^{\alpha(n)}$ with
$\beta = \alpha + 1$. These exponents $\alpha$ and $\beta$ 
are sample-independent and depend only on the nature of the
dominant disorder. We provide asymptotic theoretical analysis of
$\alpha$ and $\beta$ (for various types of underlying disorder) in the
high and the low density regime and for near and far impurities. We
have then verified our analytical results with direct numerical
calculations based on the full solution of the Boltzmann transport
theory in the presence of disorder scattering.

Although our work is primarily theoretical, we provide a critical
comparison with various experimental results in the literature (in 2D
n-GaAs electrons and p-GaAs holes), finding generally good agreement
between our theoretically predicted exponents $\alpha$ and $\beta$ and
low-temperature experimental findings in high-mobility 2D electron and
hole systems. In particular, a truly exciting prediction, that
$\alpha(n)$ has a maximum universal value $\alpha_{\rm max} \sim 1.7$
for all 2D systems at an intermediate carrier density value
(approximately defined by $k_F d \sim 1$), seems to be consistent with
recent (and old) experimental results from several different groups
(and for both 2D electrons and holes), as discussed in details in
Sec.~VI. 

Our theory predicts $\alpha(n)$ to vary from $\alpha=0$ in the
low-density strong-screening regime ($q_{TF} \gg 2k_F$, $n\rightarrow
0$) to $\alpha = 3/2$ in the high density weak screening regime
($q_{TF} \ll 2k_F$, $n \rightarrow \infty$) with a shallow maximum of
$\alpha \sim 1.7$ at intermediate carrier density for $k_F d \sim 1$
where $d$ is the impurity location. Although our work presented in
this article is purely theoretical describing the density dependent and
disorder-limited $T=0$ conductivity of 2D/3D carriers using the
semiclassical Boltzmann theory approach, it is worthwhile to speculate
about the prospects for the experimental observation of our asymptotic
low ($\alpha \rightarrow$; $\beta \rightarrow 1$) and high ($\alpha
\rightarrow 3/2$; $\beta \rightarrow 5/2$) density behavior of Coulomb
disorder limited 2D semiconductor transport.
These limiting exponents are theoretically universal.

We first summarize the current experimental status in the context of
our theory. For scattering by background Coulomb disorder (i.e., near
impurities with $2k_F d <1$), $\alpha(n) \approx 0.5 -0.8$ typically,
and for scattering by remote impurities ($2k_F d >1$) in the
modulation doping layer $\alpha(n) > 1-1.3$ typically with a few
atypical cases showing $\alpha(n) \sim 1.7$ ($>3/2$) around $k_F d
\sim 1$. All of these are intermediate density results in our theory
where $q_s$ ($=q_{TF}/2k_F$) and $2k_Fd$ are neither extremely large
nor extremely small. Thus the basic experimental situation is in
excellent agreement with our theory as it should be because the 2D
doped semiconductor transport (as well as graphene) is known to be
dominated by screened Coulomb disorder with near or far charged
impurities being the dominant scattering mechanism depending on the
sample and carrier density.

First, we discuss the high-density situation which is theoretically
more straightforward and where the Boltzmann theory is almost
exact. As carrier density increases, the semiclassical 
Boltzmann theory becomes increasingly more valid for Coulomb disorder
limited transport properties since the conductivity itself and
consequently $k_F l$ increases, thus making the system progressively
more metallic. In addition, increasing density decreases the metallic
$r_s$-parameters, the dimensionless Wigner-Seitz radius, given by $r_s
= me^2/(\kappa \hbar^2 \sqrt{\pi n})$ for 2D semiconductor
systems. Since $r_s \sim n^{-1/2}$, at high carrier density (e.g.,
$r_s = 0.5$, 2.5 for 2D n-GaAs, p-GaAs respectively at $n=10^{12}$
cm$^{-2}$) $r_s$ is small, making our theory using RPA screened
effective Coulomb disorder to be systematically more valid at higher
carrier density as RPA becomes exact at low $r_s$. Thus, it appears
that the ideal applicability of our theory is in obtaining the
high-density 2D system.

This is indeed true except that new physical
(rather than theoretical) complications arise making it problematic
for a direct comparison between 
our theory and experiment on 2D systems at high carrier density. Two
new elements of physics come into play at high carrier density, both
contributing to the suppression (enhancement) of mobility (scattering
rate): Intersubband scattering becomes important as the Fermi level
moves up and comes close to (or crosses over into) the higher confined
subbands of the quasi-2D quantum well, thus opening up a new
scattering channel; and, short-range scattering at the interface and
by alloy disorder (in AlGaAs) becomes important as the self-consistent
electric field created by the electrons themselves pushes the carriers
close to the interface at high carrier density. Both of these physical
effects eventually suppress the monotonic growth of $\mu (n)$ and
$\sigma(n)$ with increasing density, and eventually $\mu(n)$ starts
decreasing with increasing carrier density at high enough density (for
$n > 3 \times 10^{11}$ cm$^{-2}$ in GaAs-AlGaAs systems) instead of
continuing as $\mu(n) \sim n^{3/2}$ as it would
in the high-density regime if Coulomb
disorder is the only dominant scattering mechanism with no other
complications. Since at high density $\tau^{-1}(n) \sim n^{\alpha}$
with $\alpha > 1$ for Coulomb disorder, eventually at some (nonuniversal
sample dependent) high density, Coulomb disorder becomes insignificant
compared with the short-range scattering effects. Obviously, the
density at which this happens is non-universal and depends on all the
details of each sample. But in all 2D systems and samples, eventually,
when Coulomb disorder induced scattering rate is sufficiently small, a
high density regime is reached where the
mobility stops increasing (and even starts decreasing). In Si MOSFETs
\cite{ten} this effect is very strong already around $\sim 10^{12}$
cm$^{-2}$ because of considerable surface roughness scattering at the
Si-SiO$_2$ interface, and $\mu(n)$ decreases with increasing
$n$ at higher density. Even in high-mobility GaAs systems, $\mu(n)$
saturates 
(i.e., $\alpha=0$) and eventually starts decreasing at a non-universal
density around $n \agt 3\times 10^{11}$ cm$^{-2}$. Boltzmann transport
theory can be easily generalized to incorporate inter-subband
scattering and surface scattering, but the physics is nonuniversal and
beyond the scope of our current work.

The low-density ($n \rightarrow 0$) situation is fundamentally
inaccessible to our Boltzmann transport theory since all doped
semiconductor systems (3D or 2D) eventually undergo an effective
metal-insulator transition at (a nonuniversal) low (critical) density
with the semiclassical Boltzmann theory eventually becoming invalid as
a matter of principle at a sufficient low sample-dependent carrier
density. In 3D, this transition may be a true Anderson localization
transition as $k_F l \rightarrow 1$ with decreasing density making the
Boltzmann theory inapplicable. In 2D semiconductors, which are the
systems of our main interest, the observed metal-insulator transition
at a nonuniversal sample dependent critical density $n_c$ is likely to
be a crossover phenomenon \cite{five,nine} 
since the scaling theory of Anderson localization predicts 2D to be
the critical dimension with no localization transition. There is
considerable experimental support for the observed low-density 2D
metal-insulator transition to be a density inhomogeneity-driven
percolation transition at a nonuniversal critical density $n_c$ where
charged impurity induced Coulomb disorder drives the system into an
inhomogeneous collection of ``puddles'' with a mountain and lake
potential landscape where semiclassical metallic transport becomes
impossible for $n < n_c$ with $n_c$ being the disorder-dependent
percolation transition density
\cite{twenty,twentyone,leturcq,tracy2009,adam2}. The critical density $n_c$ 
depends crucially on the sample quality and is typically 
below $10^9$ cm$^{-2}$ in high-mobility GaAs systems --- in Si
MOSFETs, where disorder is very strong, $n_c \approx 10^{11}$
cm$^{-2}$, and this is why we have left out 2D Si systems from our
consideration in this work although our basic theory applies well to
2D Si-based systems for $n > 10^{11}$ cm$^{-2}$.

Our semiclassical Boltzmann theory works for $n \gg n_c$, but for $n
\rightarrow n_c$, one must include the inhomogeneous puddle formation
and the associated percolation transition even in the semiclassical
theory \cite{hwang}. This is the reason our theory fails for real 2D
systems in the $n \rightarrow 0$ limit unless the level of disorder is
extremely low (even then our Boltzmann theory is valid only for $n \gg
n_c$ and the $n \rightarrow 0$ limit is fundamentally inaccessible). 

In spite of this fundamental difficulty in accessing the low-density
(i.e. $n \rightarrow 0$) limit using our theory directly, it turns out
that we can approximately include the effect of a semiclassical
percolation transition in our theory by an indirect technique in the
density regime $n > n_c$ above the effective metal-insulator
transition (but still in a reasonably low density regime for
high-quality samples where $n_c$ can be very low). Since the
percolation picture essentially eliminates a certain fraction of the
carriers from being metallic, we can assume that the effective
conductivity (mobility) for $n \gg n_c$ is given by the same exponent
$\beta$ ($\alpha$) calculated in our theory with the only caveat that
$\beta$ ($\alpha$) is now the exponent only for the actual ``metallic''
free mobile carrier fraction of the whole
system. We can then write:
\begin{equation}
\mu = A (n-n_c)^{\alpha} = B n^{\alpha'},
\label{eq:eq68}
\end{equation}
where $\alpha(n)$ is the real exponent we calculate theoretically from
the Boltzmann theory and $\alpha'(n)$ is the effective (apparent)
exponent obtained experimentally from
\begin{equation}
\alpha'(n) = \frac{d \ln \mu(n)}{d \ln n}
\label{eq:eq69}
\end{equation}
by using the actual data for $\mu(n)$ without taking into account any
complications arising from the existence of $n_c$. Using
Eqs.~(\ref{eq:eq68}) and (\ref{eq:eq69}), we immediately get the
following relationship connecting the effective exponent $\alpha'$
with the real exponent $\alpha$ for $n \gg n_c$
\begin{equation}
\alpha' = \alpha (1-n_c/n)^{-1},
\label{eq:eq70}
\end{equation}
for $n \gg n_c$. We note that $\alpha' \approx 2 \alpha$ for $n =
2n_c$. Eq.~(\ref{eq:eq70}) is valid within logarithmic accuracy, and
connects the measured low-density ($n$ small, but $n \gg n_c$ being
still valid) exponent $\alpha'(n)$ with the real Boltzmann theory
exponent (as in Table I) $\alpha(n)$. Since $n_c$ can be measured
experimentally by checking where $\sigma(n)$ vanishes, and defining
$n_c$ to be $\sigma(n_c) = 0$ at $T=0$, we then immediately see that
the measured low-density effective exponent $\alpha'(n) > \alpha(n)$
always, and in fact, $\alpha'(n) \rightarrow \infty$ as $n \rightarrow
n_c^+$. Thus, we conclude that the experimentally measured mobility
exponent will eventually start increasing as $n$ decreases approaching
$n_c$. For $n \gg n_c$, we get $\alpha' \approx \alpha$, but the
leading correction to $\alpha$ goes as $\alpha' = \alpha (1+n_c/n)$
for $n \gg n_c$. We note that the existence of $n_c$ enhances $\alpha$
over its nominal value of Table I even for $n \gg n_c$!

We have checked existing experimental results in the literature (to the
extent $n_c$, $\alpha'$ etc. are known experimentally), finding that
our prediction of Eq.~(\ref{eq:eq70}) seems to apply quite well. We
note that since the pristine calculated $\alpha(n)$ decreases with
decreasing $n$ (see Figs.~\ref{fig:one}--\ref{fig:eight}) for {\it
  all} models of Coulomb disorder at low carrier density,
Eq.~(\ref{eq:eq70}) predicts that the apparent exponent $\alpha'$
would be close to $\alpha$ for $n \gg n_c$, but would then manifest a
minimum at a density $n_0$ ($>n_c$) defined by the equation:
\begin{equation}
\frac{d\alpha}{dn} = \frac{\alpha (n_c/n)}{(n-n_c)}
\end{equation}
at $n=n_0>n_c$. For $n <n_0$, $\alpha'$ will increase as $n$
decreases (with $\alpha'$ being a minimum at $n=n_0$), eventually 
diverging as $(1-n_c/n)^{-1}$ as $n \rightarrow n_c^+$. We find this
behavior to be qualitatively consistent with all the existing data in
the literature 
although a precise quantitative comparison necessitates more low
temperature data showing $\mu(n)$ all the way down to $n =n_c$ where
$\mu(n) = 0$. Careful measurements of $\sigma(n)$ close to $n_c$ are
lacking in the literature for us to form a definitive conclusion on
this matter at this stage.

We conclude by pointing out that very close to the percolation
transition, where $n-n_c \ll n_c$, i.e. $n \ll 2 n_c$, we expect the
critical behavior of 2D percolation transition to possibly come into
play, where the conductivity may have a completely different universal
2D percolation exponent $\delta$ (totally distinct from $\alpha$ or
$\alpha'$) which has nothing to do with our Boltzmann theory:
\begin{equation}
\sigma \sim (n-n_c)^{\delta}
\end{equation}
for $(n-n_c) \ll n_c$. This percolation critical exponent $\delta$ for
$\sigma(n)$
can only manifest itself very close to $n_c$ (i.e. $n \ll 2n_c$), and
for $ n \agt 2 n_c$ we believe that our effective theory predicts an
effective conductivity exponent $\beta' = \alpha' + 1$ given by
\begin{equation}
\beta' \approx \frac{\alpha}{1-n_c/n} + 1 = \frac{\beta -
  n_c/n}{1-n_c/n},
\label{eq:eq72}
\end{equation}
for $n \agt 2 n_c$.
Again, as for $\alpha'$, $\beta' \approx \beta = \alpha + 1$ for $n
\gg n_c$. The conductivity exponents $\delta$ and $\beta'$ arise from
completely different physics (from percolation critical theory near
$n_c$ and Boltzmann theory far above $n_c$, respectively) and have
nothing to do with each other. How the transition 
or crossover occurs even within the semiclassical theory from $\beta'$
(for $n \agt 2n_c$) to $\delta$ for ($n \ll 2n_c$) is a very
interesting question which is beyond the scope of the current work. In
our current work, we have, however, solved [Eqs.~(\ref{eq:eq68}) --
(\ref{eq:eq72}) above] the problem of the crossover from the effective
exponent $\alpha'$ (or $\beta'$) for $n \agt 2 n_c$ to the true
Boltzmann exponent $\alpha$ (or $\beta$) for $n \gg 2n_c$.

We give one possible experimental example for the observed crossover
from the Boltzmann exponent $\alpha$ (for $ n \gg n_c$) to $\alpha'$
(for $n \agt n_c$) arising in Jiang {\it et al.} \cite{jiang}. In this
work \cite{jiang}, the conductivity was measured in a high-mobility
modulation doped 2D GaAs electron system, finding the effective
mobility exponent $\alpha \approx 0.9-1.1$ in the high density range
($n \sim 5 \times 10^{10} - 3 \times 10^{11}$ cm$^{-2}$) for $d=350 -
750$ \AA. This range of $n$ and $d$ converts to $2k_F d = 1-5$ and
$q_s \approx 1$, which is the intermediate density range for all our
Coulomb disorder mechanisms in Table I. Given that the measured
mobility in Ref.~[\onlinecite{jiang}] was relatively modest, $\mu \sim
10^5 - 10^6$ cm$^2$/Vs, it is reasonable to expect, based on our
numerical mobility calculations, for both remote and background
Coulomb scattering to be equivalently effective, leading to $\alpha
\sim 0.9-1.2$ according to our numerical results of Sec.~IV. Thus, the
``high-density'' exponent ($\alpha \sim 1$) in the Jiang sample agrees
well with our theory. The interesting point to note in the current
context is that Jiang {\it et al.} \cite{jiang} found a large increase
of the measured exponent $\alpha$ as $n$ decreases, which is in
apparent disagreement with our analytical theory which always gives
$\alpha(n)$ decreasing with decreasing $n$. Although other
possibilities (e.g., multiscattering) cannot be ruled out \cite{gold},
we believe that the Jiang {\it et al.} data indicate a crossover from
$\alpha$ to $\alpha'$ as $n$ decreases. In particular, the measured
mobility exponent increased from $\alpha \sim 1$ to $\alpha \sim 4$ as
$n$ decreased from $\sim 10^{11}$ cm$^{-2}$ to $\sim 2 \times 10^{11}$
cm$^{-2}$ (with also a factor of 50 decrease in mobility). Assuming
$n_c \sim 1.5 \times 10^{11}$ cm$^{-2}$ (which is consistent with the
data), we get [from Eq.~(\ref{eq:eq70})], $\alpha' \approx 4 \alpha \approx
4$ at $n \approx 2 \times 10^{10}$ cm$^{-2}$, which is in excellent
agreement with the data \cite{jiang}. We therefore believe that the
Jiang {\it et al.} experiment manifests our predicted crossover
behavior from $\alpha(n)$ to $\alpha'(n)$ as $n$ approaches $n_c$ from
above.

We note that $k_F l \approx 1$ in the Jiang {\it et al.} sample for $n
\approx 3 \times 10^{10}$ cm$^{-2}$ (where $\mu \approx 10^5$
cm$^2$/Vs seems to have been reached starting from $\mu \sim 10^6$
cm$^2$/Vs for $n \sim 3 \times 10^{11}$ cm$^{-2}$), and thus the
identification of $n_c \approx 1.5 \times 10^{10}$ cm$^{-2}$ is
meaningful. Clearly, our simple Drude-Boltzmann theory applies for $n
\gg 3 \times 10^{10}$ cm$^{-2}$, but not for $n < 3 \times 10^{10}$
cm$^{-2}$ where $k_F l \sim 1$. What is encouraging is that the simple
modification of the theory introducing the crossover exponent $\alpha'
= \alpha (1-n/n_c)^{-1}$ seems to describe the experimentally observed
density scaling exponent of the observed experimental mobility at low
carrier densities.

We conclude by mentioning that direct numerical percolation
calculations indicate $\delta \approx 1.32$ in 2D systems, which is
unfortunately too close to our calculated value of $\beta$ in the
low-density Coulomb disorder-dominated Boltzmann theory, where (see
Figs.~\ref{fig:one} -- \ref{fig:six} and \ref{fig:eight}) the
low-density ($n \alt 10^{10}$ cm$^{-2}$) $\alpha$-value is $\alpha
\approx 0.3-0.5$ implying low-density $\beta \approx 1.3-1.5$. Since
high-mobility 2D GaAs samples typically have $n_c \approx 10^9$
cm$^{-2}$, it is unclear whether the existing conductivity exponent
measurements for the putative metal-insulator transition
\cite{twenty,twentyone,leturcq,tracy2009} for 
the 2D density-driven metal-insulator transition really is obtaining
$\delta$ or is just a measurement of our calculated $\beta=\alpha + 1$
which at low values of $n$ would be rather close to the experimentally
measured percolation exponent of $1.3-1.5$. The current experiments do
not really quantitatively 
measure $\sigma(n)$ in the $n < 2 n_c$ regime necessary for obtaining
$\delta$, and we feel that much more work will be needed to establish
the nature of the $\sigma(n \rightarrow n_c) \rightarrow 0$ transition
observed in the laboratory. It is possible, even likely, that the
existing measurements have only measured the low density (but still $n
\gg n_c$) value of our calculated (non-critical) Boltzmann exponent
$\beta \approx 1.3-1.5$ in n- and p- 2D GaAs systems.

In conclusion, we have developed a comprehensive theory for the
universal density scaling of the low-temperature transport properties
of 2D and 3D doped semiconductors and graphene, concentrating on the
role of background Coulomb disorder and obtaining, both theoretically
and numerically,  the power law exponents for the density-dependent
mobility and conductivity within the Boltzmann transport theory.

\section*{acknowledgments}
This work was supported by Microsoft Q, JQI-NSF-PFC, DARPA QuEST,
LPS-CMTC, and US-ONR.


\end{document}